\documentclass[preprint,showpacs,preprintnumbers,amsmath,amssymb,nofootinbib,axodraw]{revtex4}
\usepackage{graphicx}% Include figure files
\usepackage{dcolumn}% Align table columns on decimal point
\usepackage{bm}% bold math
\usepackage{axodraw}
\usepackage{color}

\def\cO#1{{\cal{O}}\left(#1\right)}

\newcommand{\beq}{\begin{equation}}
\newcommand{\eeq}{\end{equation}}
\newcommand{\beqa}{\begin{eqnarray}}
\newcommand{\eeqa}{\end{eqnarray}}
\newcommand{\nn}{\nonumber\\}
\newcommand{\e}{\epsilon}
\def\Ll{{\rm L}}

\def\li{{\rm Li_2}}
\def\cG{c_\Gamma}
\def\pu{p_1}
\def\pd{p_2}
\def\pt{p_3}
\def\pq{p_4}
\def\qu{q_1}
\def\qd{q_2}
\def\qt{q_3}
\def\qq{q_4}
\def\sud{s_{12}}

\def\sdt{s_{23}}

\def\slash#1{#1 \hskip-0.45em /}

\begin{document}

\preprint{FERMILAB-PUB-05-354-T}

\title{Semi-Numerical Evaluation of One-Loop Corrections}

\author{R.~K.~Ellis}
\email{ellis@fnal.gov}
\author{W~T.~Giele}
\email{giele@fnal.gov}
\author{G.~Zanderighi}
\email{zanderi@fnal.gov}
\affiliation{ Fermilab, Batavia, IL 60510, USA }

\date{\today}
\begin{abstract}
  We present a semi-numerical algorithm to calculate one-loop virtual
  corrections to scattering amplitudes.  The divergences of the loop
  amplitudes are regulated using dimensional regularization. We only 
  treat in detail the case of amplitudes with up to five external legs 
  and massless internal lines, although we believe the method to be 
  more generally applicable. Tensor integrals are reduced to generalized scalar
  integrals, which in turn are reduced to a set of known basis
  integrals using recursion relations.  The reduction algorithm is
  modified near exceptional configurations to ensure numerical
  stability. To test the procedure we apply these techniques to
  one-loop corrections to the Higgs to four quark process for which
  analytic results have recently become available.

\end{abstract}

\maketitle
\section{Introduction}
In order to make reliable theoretical estimates of scattering
amplitudes it is important to calculate next-to-leading order (NLO)
corrections.  Such calculations will give a reliable estimate of the
cross section normalization.  In addition, they can also give results
for kinematic distributions.  The latter are important for
understanding backgrounds to new physics searches in greater detail.

While most of the two-body final states at hadron colliders have been
calculated at NLO and implemented in flexible numerical
programs~\cite{Baer:1990ra,Mangano:1991jk,Bailey:1992br,Ellis:1992en,
  Giele:1993dj,Campbell:1999ah,Binoth:1999qq,Dixon:1999di}, results
for three- or four-body final states are more recent and as yet quite
limited
\cite{Kilgore:1996sq,Campbell:2002tg,Beenakker:2002nc,Nagy:2003tz,Campbell:2003hd,Denner:2005es}.
One of the difficulties is the complexity of the analytic calculations
required to evaluate the virtual contributions.  The computation of
leading-order (LO) amplitudes has been automated using a variety of
methods~\cite{Berends:1987me,Murayama:1992gi,Stelzer:1994ta,
  Caravaglios:1995cd}.  Explicit algorithms for numerical evaluations
of LO cross sections have been developed
\cite{Pukhov:1999gg,Yuasa:1999rg,Kanaki:2000ey,Krauss:2001iv,
  Mangano:2002ea,Maltoni:2002qb} and are of great value to
experimenters.  In contrast, no similar development has occurred for
the evaluation of NLO amplitudes.
Because of recent progress in the understanding of algorithms to
evaluate one-loop
integrals~\cite{Ferroglia:2002mz,Denner:2002ii,Duplancic:2003tv,Giele:2004iy,Giele:2004ub,delAguila:2004nf,vanHameren:2005ed,Binoth:2005ff,
  Denner:2005nn} one can envision a similar automation of one-loop
calculations.  This would enable us to extend NLO calculations to more
complicated final states, such as four-body final states (e.g. the
production of two pairs of massive quarks), five-body final states
(e.g. vector boson plus four jets) and beyond.

In this paper we report on early steps along this path.  
Section~\ref{general_approach} discusses a general approach to
one-loop calculations.  In section III we outline the numerical
implementation of an algorithm similar to the one developed in
ref.~\cite{Giele:2004iy} for processes with up to five external
particles.  We use the integration-by-parts method
\cite{Tkachov:1981wb,Chetyrkin:1981qh} to evaluate up to rank-five
five-point tensor integrals with massless internal lines.  In the
kinematic regions where a Gram determinant is small or vanishes, the
basic recursive algorithm discussed in Section III breaks down.
Section \ref{exceptional} shows how to treat these exceptional
regions, and gives an explicit algorithm for numerical implementation.
In particular, it discusses how to extend the basic method, (outlined
in ref.~\cite{Giele:2004ub}), to treat five-point tensor integrals in
the exceptional regions.  The results are shown to be stable close to
exceptional phase space points.  Section V provides a brief outlook.
Appendix \ref{integrals} collects results for the basis integrals.

The combined algorithms of Sections \ref{numerical} and
\ref{exceptional} have been implemented in a numerical program.  As a
first application, we have used the tensor evaluation package to
calculate the virtual corrections to Higgs plus up to four partons
using an effective Lagrangian approach~\cite{Ellis:2005qe}.

\section{Anatomy of a One-Loop Calculation}
\label{general_approach} 
\begin{figure}[t]
\includegraphics[angle=270]{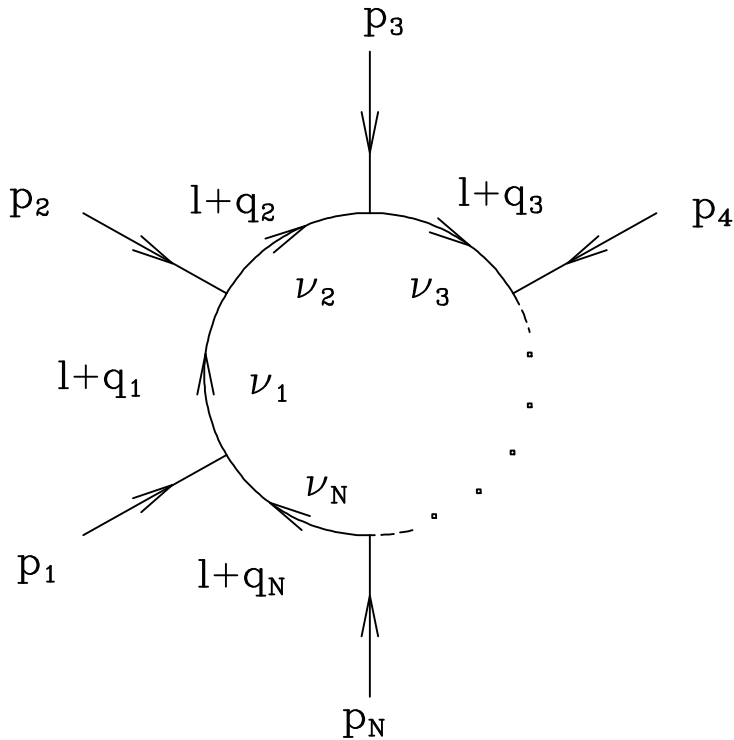}
\caption{\label{fig:generic}The generic $N$-point loop graph.}
\end{figure}

The calculation of virtual corrections to a scattering process follows
a general pattern. Each scattering amplitude ${\cal M}$ consists of a
gauge-invariant sum of $N_D$ Feynman diagrams, $\{{\cal
  A}_i\}_{i=1}^{N_D}$:
\beq
{\cal M}=\sum_{i=1}^{N_D} {\cal A}_i\ .
\eeq
Each one-loop Feynman diagram ${\cal A}$ can in turn be decomposed
into a sum of contractions between two types of tensors.  The first
tensor is the kinematic tensor $K_{\mu_1\cdots\mu_M}$ containing
couplings and the particle properties such as kinematics, spin,
polarization and color.  The other tensor is a one-loop tensor
integral $I^{\mu_1\cdots\mu_M}$ which only depends on the event
kinematics, i.e. it depends on the momenta and masses of the scattered
particles.  The explicit decomposition of each Feynman diagram is
given by
\beq
{\cal A}(\pu,\ldots, p_N;\varepsilon_1,\ldots, \varepsilon_N)=
\sum_{M=0}^N  K_{\mu_1\cdots\mu_M} 
(p_1,\ldots, p_N; \varepsilon_1,\ldots, \varepsilon_N) \; 
I^{\mu_1\cdots\mu_M}(D; q_1,\ldots,q_N)\,,
\eeq
where $p_i$ and $\varepsilon_i$ denote the momenta and polarizations of
the external particles and the tensor integral $I^{\mu_1\cdots\mu_M}$
is given by
\beq
I^{\mu_1\cdots\mu_M}(D; q_1,\ldots,q_N)\equiv
\int \frac{d\,^D l}{i \pi^{D/2}}\ \frac{l^{\mu_1}\cdots 
l^{\mu_M}}{d_1d_2\cdots d_N}\,, \quad 
d_i \equiv (l+q_i)^2\,,\quad q_i \equiv \sum_{j=1}^i p_j\,,
\eeq
and $D=4-2\e$. 
Note that we are considering the case of massless internal
propagators, although the method can be extended to include internal
masses.
The generation of the kinematic tensor $K$, is analogous to the
generation of leading-order (LO) scattering amplitudes.  That is, the
kinematic tensor is composed of LO multi-particle sources.  Many
methods have been developed to deal with the generation of LO
amplitudes, either analytic or purely algorithmic.  Such methods can
be adapted to generate the kinematic tensor.
 
The more problematic issue is the evaluation of the tensor integral.
Using the generic expansion of a tensor integral
\beqa
I^{\mu_1\cdots\mu_M}(D; q_1,\ldots,q_N)&=&
\sum_{i_1i_2\cdots i_M}^N a_{i_1i_2\cdots i_M} 
q_{i_1}^{\mu_1}q_{i_2}^{\mu_2}\cdots q_{i_M}^{\mu_M}
\nn
&+&\sum_{i_3i_4\cdots i_M}^N a_{00i_3i_4\cdots i_M} 
g^{\mu_1\mu_2}q_{i_3}^{\mu_3}q_{i_4}^{\mu_4}\cdots q_{i_M}^{\mu_M}
\nn
&+&\sum_{i_5i_6\cdots i_M}^N a_{0000i_5\cdots i_M} 
g^{\mu_1\mu_2}g^{\mu_3\mu_4}q_{i_5}^{\mu_5} q_{i_6}^{\mu_6} 
\cdots q_{i_M}^{\mu_M}
\nn
&+&\cdots\,,
\eeqa
we can decompose the tensor integral as known tensor structures and form
factor coefficients $a_{j_1j_2\cdots j_M}$.  Once we have a procedure
to calculate the form factors, we can evaluate the tensor integral
numerically using a computer program (see
also~\cite{Veltman:1979,Passarino:1978jh,vanOldenborgh:1990yc,Hahn:1998yk}).

Using integration-by-parts decomposition techniques each form factor
coefficient is identified with one generalized {\it scalar}
integral~\cite{Davydychev:1991va}, as pictured in Fig.~\ref{fig:generic},
\beq \label{scalar_integral}
I(D;\{q_1,\nu_1\},\ldots,\{q_N,\nu_N\}) \equiv 
I(D;\nu_1,\nu_2,\ldots,\nu_N)\equiv 
\int \frac{ d^D\,l}{i \pi^{D/2}} 
\; \frac{1} {d_1^{\nu_1}d_2^{\nu_2}\cdots d_N^{\nu_N}}\,,
\eeq
for which a recursive algorithm can be set up in such a manner that
all generalized scalar integrals can be {\it numerically} reduced to
the set of analytically known master integrals.  Using this method we
can construct numerical algorithms to calculate the general tensor
integral. Note that, due to the fact that the master integrals are
evaluated using analytic expressions, we do not anticipate any
significant loss of accuracy for the numerical evaluation of the
amplitude compared to a purely analytic method.

One complication is the fact that the evaluation has to be performed
within dimensional regularization. Because of this we cannot simply
calculate a complex valued tensor. Instead we calculate the complex
valued coefficients $c_i$ of a Laurent series:
\beq
a_{j_1j_2\cdots j_M}=
 \frac{c_{-2}}{\e^2}+\frac{c_{-1}}{\e}+c_0+c_1 {\e} + {\cal O}(\e^2)\,,
\eeq
as was advocated in~ref.~\cite{vanHameren:2005ed}.  As the tensor
integral contains terms of order $1/\e^2$ and $1/\e$ we need to
specify a consistent scheme for evaluating the kinematic tensor up to
potential terms of order $\e^2$.

All the elements for a fully automated evaluation of one-loop virtual
amplitudes are now in place, although they are not yet implemented in
a single procedure. Indeed, the method we used relies on the
generation of the graphs using Qgraf~\cite{Nogueira:1991ex}, and the
simplification of terms proportional to the square of the
loop-momentum using the algebraic manipulation program
Form~\cite{Vermaseren:2000nd}. This latter step is a necessary
prerequisite to evaluate the kinematic tensor. Were terms proportional
to $g^{\mu \nu}$ to remain in the kinematic tensor, they would lead to
ambiguities in performing the contraction with the tensor
integral.~\footnote{By eliminating all metric tensors from the
  kinematic matrix we ensure that the kinematic tensor is composed
  entirely of external vectors which are purely four dimensional.}

For amplitudes involving more than one quark line one can encounter
terms of the form $ \bar u(p_1)\dots \slash{l}\dots u(p_2)\, \bar
u(p_3)\dots \slash{l} \dots u(p_4)$, (where $\slash{l}=\gamma_\mu l^\mu$), 
which can not be simplified. For
the simple case $H \to q \bar q q' \bar q'$ (and $H \to q \bar q q
\bar q$) we performed the square analytically, thus removing all
spinor lines. After elimination of the spinors, the dangerous terms
quadratic in $l$ can always be removed.

\section{Numerical implementation of the algorithm}
\label{numerical}
In this section we will outline the numerical implementation of the
evaluation of tensor integrals
$I^{\mu_1\mu_2\cdots\mu_M}(D;q_1,q_2,\ldots,q_N)$ in terms of a
Laurent series. The first step is to reduce the form factor
coefficients $a_{j_1j_2\cdots j_M}$ to the generalized scalar
integrals $I(D;\nu_1,\nu_2,\ldots,\nu_N)$. Next, the generalized
scalar integrals have to be reduced to the set of master integrals.
Finally, the master integrals have to be evaluated.

\subsection{The tensor integral decomposition}
The tensor integral decomposition is performed using the method of
Davydychev~\cite{Davydychev:1991va}, which is derived using the
integration-by-parts method.  We introduce the notation for the
generalized scalar integrals, eq.~(\ref{scalar_integral}),
\beq
I(D;\{\nu_k\}_{k=1}^N) \equiv  I(D;\nu_1,\nu_2,\ldots,\nu_N) \, .
\eeq
The general form for the tensor integrals with $N$ external legs and
$M$ free indices can be expressed in terms of these scalar
integrals
\begin{eqnarray} \label{Davyd}
I_{\mu_1\ldots\mu_M}(D;\{q_1,\nu_1\}\ldots,\{q_N,\nu_N\}) &=& 
\sum_{\substack{
\lambda,\kappa_1,\kappa_2,\ldots, \kappa_N  \ge 0\\
2 \lambda+\sum_i \kappa_i =M}
}
\left(-{\frac{1}{2}}\right)^{\lambda} \;  
\{ [g]^\lambda [q_1]^{\kappa_1} \ldots 
[q_N]^{\kappa_N} \}_{\mu_1\ldots \mu_M}
\\
&\times& (\nu_1)_{\kappa_1} \ldots (\nu_N)_{\kappa_N} \; 
I(D+2(M-\lambda);\nu_1+\kappa_1,\ldots,\nu_N+\kappa_N)\nonumber \,,
\end{eqnarray}
where 
\beq
(\nu)_\kappa=\frac{\Gamma(\nu+\kappa)}{\Gamma(\nu)}\,, 
\eeq
is the Pochhammer symbol.  The structure $\{ [g]^\lambda
[q_1]^{\kappa_1} \ldots [q_N]^{\kappa_N} \}_{\mu_1\ldots \mu_M}$ is a
tensor which is completely symmetric in the indices $\mu_1, \ldots,
\mu_M$ and constructed from $\lambda$ metric tensors $g$, $\kappa_1$
momenta $p_1$, $\kappa_2$ momenta $p_2$, etc. Once we have a method 
to evaluate the generalized scalar integrals,
eq.~(\ref{Davyd}) can be used to construct the tensor integral.

Thus, for example, the rank-1 tensor integral with $N$ external legs can be
decomposed as
\beqa
I^{\mu_1}(D; q_1, \dots, q_N)&=&
\sum_{i_1=1}^N I(D+2;\{1+\delta_{i_1k}\}_{k=1}^N)\ q_{i_1}^{\mu_{1}} \nn 
&=&
I(D+2;2,1,1,\ldots,1)\ q_1^{\mu_1}+I(D+2;1,2,1,\ldots,1)\ q_2^{\mu_1}\nn
&+&\cdots + I(D+2;1,1,1,\ldots,2)\ q_N^{\mu_1}\,.
\eeqa
Similarly, a rank-2 tensor integral is given by
\beqa
I^{\mu_1\mu_2}(D; q_1, \dots, q_N)&=&
-\frac{1}{2} \; I(D+2;\{1\}_{k=1}^N)\ g^{\mu_1\mu_2} \nn
&&+\sum_{P\{\mu_1,\mu_2\}}^{2!} \sum_{i_1\leq i_2}^N 
I(D+4;\{1+\delta_{i_1k}+\delta_{i_2k}\}_{k=1}^N)\ 
q_{i_1}^{\mu_{1}}q_{i_2}^{\mu_{2}} \nn 
&=& -\frac{1}{2}I(D+2;1,1,1,\ldots,1)\ g^{\mu_1\mu_2} \nn
&&+2 \; I(D+4;3,1,1,\ldots,1)\ q_1^{\mu_1}q_1^{\mu_2} \nn 
&&+\phantom{2}\;I(D+4;2,2,1,\ldots,1)\ 
\left(q_1^{\mu_1}q_2^{\mu_2}+q_1^{\mu_2}q_2^{\mu_1}\right)\nn
&&+\cdots\,,
\eeqa
where the first sum runs over the permutations of the indices
$P\{\mu_1,\mu_2\}$.

\subsection{Recursion scheme for scalar integrals}
We will outline and improve the recursive scheme proposed in ref.
\cite{Giele:2004iy}, presenting explicit formulas for the case of five
or less external particles. Using the basic equation of the
integration-by-parts method, an identity which is valid for arbitrary
values of the parameters $y_i$ can be derived,
\beq
\int\frac{d^Dl}{i \pi^{D/2}}\frac{\partial }{\partial l^\mu }\left(
\frac{\left(\sum_{i=1}^N y_i\right)l^{\mu}
+\left(\sum_{i=1}^N y_iq_i^{\mu}\right)}
{d_1^{\nu_1}d_2^{\nu_2}\cdots d_N^{\nu_N}}\right) = 0\,.
\eeq
Differentiating we obtain the base identity
\beqa \label{baseidentity}
\lefteqn{\sum_{j=1}^N\left(\sum_{i=1}^N S_{ji}y_i\right)\nu_j 
I(D;\{\nu_k+\delta_{kj}\}_{k=1}^N)=} \nn
&&-\sum_{i=1}^N y_i I(D-2;\{\nu_k-\delta_{ki}\}_{k=1}^N)
-\left(D-1-\sum_{j=1}^N\nu_j\right)\left(\sum_{i=1}^N y_i\right)
I(D;\{\nu_k\}_{k=1}^N)\,,
\eeqa
where $S$ is a kinematic matrix which, for massless internal particles,
takes the form 
\beq \label{Sdef}
S_{ij}\equiv \left(q_i-q_j\right)^2.
\eeq
The dimensional shift identity
\beq 
\label{eq:dimshift}
I(D-2;\{\nu_k\}_{k=1}^N)=-\sum_{i=1}^N\nu_i 
I(D;\{\nu_k+\delta_{ik}\}_{k=1}^N)\,,
\eeq 
was used to obtain the base identity.

The integrals are characterized by the value of $n$, which is the 
naive degree of ultraviolet divergence
\beq
\label{eq:ndef}
n= \Big( \frac{D}{2}-\sigma\Big) \Big|_{\epsilon=0}\, .
\eeq

\begin{figure}
\begin{center}
  \unitlength 1.pt\SetScale{1.0}
\begin{picture}(180,180)(0,0) 
%------------------------------%
  \SetOffset(0,0)
  \Line(-4,10)(0,14)
  \Line(-4,10)(0,6)
  \Line(4,10)(0,14)
  \Line(4,10)(0,6)
  \Line(0,10)(0,180)
  \Line(0,10)(180,10)
  \Text(0,0)[c]{2}
  \Text(40,0)[c]{3}
  \Text(80,0)[c]{4}
  \Text(100,-10)[c]{$D/2$}
  \Text(120,0)[c]{5}
  \Text(160,0)[c]{6}
  \Text(-8,10)[c]{3}
  \Text(-8,50)[c]{4}
  \Text(-8,90)[c]{5}
  \Text(-16,110)[c]{$\sigma$}
  \Text(-8,130)[c]{6}
  \Text(-8,170)[c]{7}
  \Line(0,10)(160,170)
  \Line(40,10)(160,130)
  \Text(180,170)[c]{n$=-1$}
  \Text(180,130)[c]{n$=0$}
  \SetColor{Green}
  \Vertex(120,130){3}
  \SetWidth{1.2}
  \DashArrowLine(120,130)(80,90){5}
  \DashArrowLine(120,130)(80,50){5}
  \SetColor{Blue}
  \Vertex(80,50){3}
  \SetWidth{1.2}
  \ArrowLine(80,50)(40,10)
  \ArrowLine(80,50)(40,50)
%------------------------------%
\end{picture}
\end{center}
\caption{\label{fig:triangle3}Reduction scheme for triangle diagrams
with three off-shell legs.}
\end{figure}
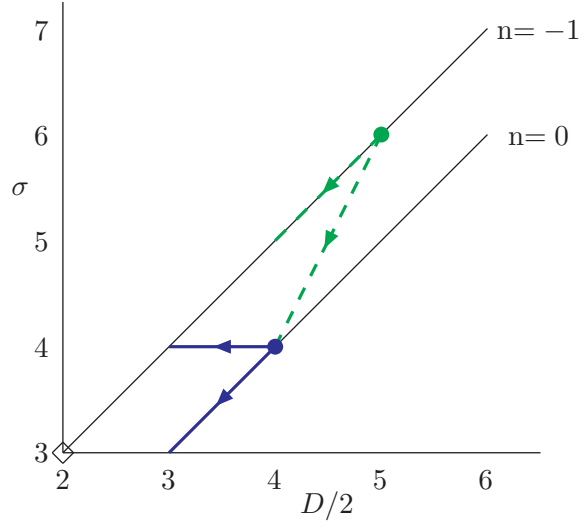

It is also helpful to introduce some further notation 
\beq
\label{eq:bidef}
\sigma\equiv\sum_{i=1}^N\nu_i;\ \quad b_i\equiv\sum_{j=1}^N S_{ij}^{-1};\ 
\quad B\equiv\sum_{i=1}^N b_i=\sum_{i,j=1}^N S_{ij}^{-1}\,.
\eeq
The basic recursion relation is obtained by solving $\sum_i S_{ji} y_i
= \delta_{lj}$ in eq.~\eqref{baseidentity} and shifting $\nu_l$ by one unit.
The result is,
\begin{eqnarray}
\label{eq:recursion1}
\lefteqn{(\nu_l-1) I(D;\{\nu_k\}_{k=1}^N)}\nonumber \\
&=&-\sum_{i=1}^N S_{li}^{-1} I(D-2;\{\nu_k-\delta_{ik}-\delta_{lk}\}_{k=1}^N)
-b_l\left(D-\sigma\right)I(D;\{\nu_k-\delta_{lk}\}_{k=1}^N). 
\end{eqnarray}
Figs.~\ref{fig:triangle3},~\ref{fig:box} and~\ref{fig:pent} illustrate
the action of the recursion relations in the plane defined by
$D/2$ and$\sigma$. The location of the basis integrals is indicated by a
diamond. The recursion relation Eq.~\eqref{eq:recursion1}
either reduces both the dimension $D$ and the value of $\sigma$ by 2, or
it keeps $D$ fixed and reduces $\sigma$ by 1. Its action is indicated by
the red (dotted) lines in Figs.~\ref{fig:box} and \ref{fig:pent}.

%OLDThis relation either reduces both the dimension $D$ and the value of
%OLD$\sigma$ by 2, or it keeps $D$ fixed and reduces $\sigma$ by 1.  Its
%OLDaction is indicated by the red (dotted) lines in Figs.~\ref{fig:box}
%OLDand \ref{fig:pent}. The location of the basis integrals is shown by a
%OLDdiamond in Figs.~\ref{fig:triangle3},~\ref{fig:box}
%OLDand~\ref{fig:pent}.

By performing the shift $\{\nu_k\}_{k=1}^N \to \{\nu_k
+\delta_{lk}\}_{k=1}^N$, summing eq.~(\ref{eq:recursion1}) over the
index $l$, and using eq.~\eqref{eq:dimshift} we derive the standard
recursion relation~\cite{Binoth:1999sp},
\begin{equation}
\label{eq:recursion2}
\left(D-1-\sigma\right)\,B\,I(D;\{\nu_k\}_{k=1}^N)=I(D-2;\{\nu_k\}_{k=1}^N)
-\sum_{i=1}^Nb_iI(D-2;\{\nu_k-\delta_{ik}\}_{k=1}^N).
\end{equation}
Eq.~(\ref{eq:recursion2}) reduces $D$ by 2 and $\sigma$ by 0 or 1.
Its action is illustrated by the blue (solid) lines shown in
Figs.~\ref{fig:triangle3},~\ref{fig:box} and ~\ref{fig:pent}.
 
For the case of pentagon diagrams with all $\nu_k=1$ and $D=4-2\e$, we
will use the following recursion relation to arrive at the basis
pentagon in $D=6-2\e$ dimensions, plus boxes
\begin{equation}
\label{eq:recursion2a}
I(D;\{\nu_k\}_{k=1}^N)=\left(D+1-\sigma\right)\,B\,I(D+2;\{\nu_k\}_{k=1}^N)
+\sum_{i=1}^N b_iI(D;\{\nu_k-\delta_{ik}\}_{k=1}^N).
\end{equation}
Its action is illustrated by the magenta (solid grey) line in
Fig.~\ref{fig:pent}.

By combining eqs.~(\ref{eq:recursion1}) and (\ref{eq:recursion2}) we
can eliminate the term with a prefactor depending on the dimension
$D$, to yield a fourth recursion relation,
\begin{eqnarray}
\label{eq:recursion3}
\lefteqn{(\nu_l-1) I(D;\{\nu_k\}_{k=1}^N)}\nonumber \\
&=&-\frac{b_l}{B}I(D-2;\{\nu_k-\delta_{lk}\}_{k=1}^N)
+\sum_{i=1}^N\left(\frac{b_lb_i}{B}-S_{li}^{-1}\right) 
I(D-2;\{\nu_k-\delta_{ik}-\delta_{lk}\}_{k=1}^N).\nonumber \\
\end{eqnarray}
This identity is particularly efficient since it reduces $D$ by 2 and
$\sigma$ by either 1 or 2 units. Its action is indicated in
Figs.~\ref{fig:triangle3},~\ref{fig:box} and \ref{fig:pent} by the
green (dashed) lines.

Note that for the case $N=3$ in Fig.~\ref{fig:triangle3} we assume
that $\det(S)\neq 0$, which is only the case if all three external
momenta are off-shell. Triangles with only one off-shell leg are basis
integrals. The case of triangle integrals with two off-shell legs will
be treated in the following section.

\begin{figure}
\label{fig:path45}
\begin{minipage}[t]{2.5in}
\begin{center}
  \unitlength 1.pt\SetScale{1.0}
\begin{picture}(180,180)(0,0) 
%------------------------------%
  \SetOffset(0,0)

  \Line(-4,10)(0,14)
  \Line(-4,10)(0,6)
  \Line(4,10)(0,14)
  \Line(4,10)(0,6)

  \Line(0,10)(0,180)
  \Line(0,10)(180,10)
  \Text(0,0)[c]{2}
  \Text(40,0)[c]{3}
  \Text(80,0)[c]{4}
  \Text(100,-10)[c]{$D/2$}
  \Text(120,0)[c]{5}
  \Text(160,0)[c]{6}
  \Text(-8,10)[c]{4}
  \Text(-8,50)[c]{5}
  \Text(-8,90)[c]{6}
  \Text(-16,110)[c]{$\sigma$}
  \Text(-8,130)[c]{7}
  \Text(-8,170)[c]{8}
  \Line(0,10)(160,170)
  \Line(40,10)(160,130)
  \Line(80,10)(160,90)
  \Text(185,170)[c]{n$=-2$}
  \Text(185,130)[c]{n$=-1$}
  \Text(185,90)[c]{n$=0$}
  \SetColor{Green}
  \Vertex(160,130){3}
  \SetWidth{1.2}
  \DashArrowLine(160,130)(120,90){5}
  \DashArrowLine(160,130)(120,50){5}
  \SetColor{Blue}
  \Vertex(120,50){3}
  \SetWidth{1.2}
  \ArrowLine(120,50)(80,10)
  \ArrowLine(120,50)(80,50)
  \SetColor{Red}
  \Vertex(120,130){3}
  \SetWidth{1.2}
  \DashArrowLine(120,130)(120,90){2}
  \DashArrowLine(120,130)(80,50){2}
%------------------------------%
\end{picture}
\end{center}
\caption{\label{fig:box}Reduction scheme for box diagrams.}
\end{minipage} 
\hspace{0.5in}
\begin{minipage}[t]{2.5in}
\begin{center}
  \unitlength 1.pt\SetScale{1.0}
\begin{picture}(180,180)(0,0) 
%------------------------------%
  \SetOffset(0,0)

  \Line(36,10)(40,14)
  \Line(36,10)(40,6)
  \Line(44,10)(40,14)
  \Line(44,10)(40,6)

  \Line(0,10)(0,180)
  \Line(0,10)(180,10)
  \Text(0,0)[c]{2}
  \Text(40,0)[c]{3}
  \Text(80,0)[c]{4}
  \Text(100,-10)[c]{$D/2$}
  \Text(120,0)[c]{5}
  \Text(160,0)[c]{6}
  \Text(-8,10)[c]{5}
  \Text(-8,50)[c]{6}
  \Text(-8,90)[c]{7}
  \Text(-16,110)[c]{$\sigma$}
  \Text(-8,130)[c]{8}
  \Text(-8,170)[c]{9}
  \Line(0,10)(160,170)
  \Line(40,10)(160,130)
  \Line(80,10)(160,90)
  \Line(120,10)(160,50)
  \Text(185,170)[c]{n$=-3$}
  \Text(185,130)[c]{n$=-2$}
  \Text(185,90)[c]{n$=-1$}
  \Text(185,50)[c]{n$=0$}
  \SetColor{Green}
  \Vertex(160,130){3}
  \SetWidth{1.2}
  \DashArrowLine(160,130)(120,90){5}
  \DashArrowLine(160,130)(120,50){5}
  \SetColor{Blue}
  \Vertex(120,50){3}
  \SetWidth{1.2}
  \ArrowLine(120,50)(80,10)
  \ArrowLine(120,50)(80,50)
  \SetColor{Red}
  \Vertex(120,130){3}
  \SetWidth{1.2}
  \DashArrowLine(120,130)(120,90){2}
  \DashArrowLine(120,130)(80,50){2}
  \SetColor{Magenta}
  \Vertex(0,10){3}
  \SetWidth{1.6}
  \ArrowLine(0,10)(40,10)
%------------------------------%
\end{picture}
\end{center}
\caption{\label{fig:pent}Reduction scheme for pentagon diagrams.}
\end{minipage}
\end{figure}

\subsection{Reduction of two mass triangles}
\label{twomasstriangles}
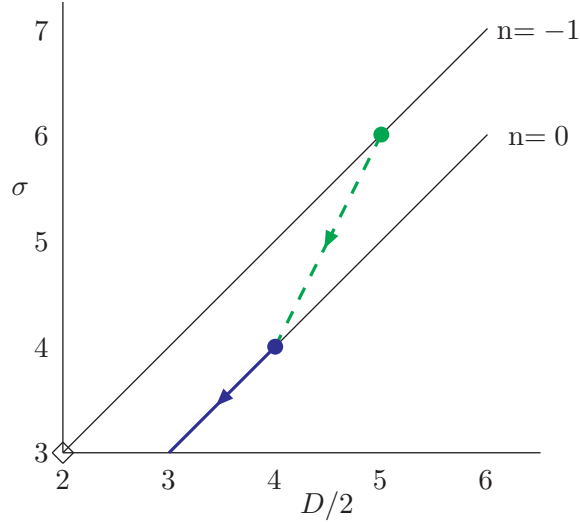
\begin{figure}
\begin{center}
  \unitlength 1.pt\SetScale{1.0}
\begin{picture}(180,180)(0,0) 
%------------------------------%
  \SetOffset(0,0)
  \Line(-4,10)(0,14)
  \Line(-4,10)(0,6)
  \Line(4,10)(0,14)
  \Line(4,10)(0,6)
  \Line(0,10)(0,180)
  \Line(0,10)(180,10)
  \Text(0,0)[c]{2}
  \Text(40,0)[c]{3}
  \Text(80,0)[c]{4}
  \Text(100,-10)[c]{$D/2$}
  \Text(120,0)[c]{5}
  \Text(160,0)[c]{6}
  \Text(-8,10)[c]{3}
  \Text(-8,50)[c]{4}
  \Text(-8,90)[c]{5}
  \Text(-16,110)[c]{$\sigma$}
  \Text(-8,130)[c]{6}
  \Text(-8,170)[c]{7}
  \Line(0,10)(160,170)
  \Line(40,10)(160,130)
  \Text(180,170)[c]{n$=-1$}
  \Text(180,130)[c]{n$=0$}
  \SetColor{Green}
  \Vertex(120,130){3}
  \SetWidth{1.2}
  \DashArrowLine(120,130)(80,50){5}
  \SetColor{Blue}
  \Vertex(80,50){3}
  \SetWidth{1.2}
  \ArrowLine(80,50)(40,10)
%------------------------------%
\end{picture}
\end{center}
\caption{\label{fig:triangle2}Reduction scheme for triangles
with two off-shell legs.}
\end{figure}

For the specific case of two mass triangles ($p_1^2,p_2^2 \ne 0, p_3^2
=0$), the matrix
\beq S=\left(
\begin{array}{ccc}
 0           & \pd^2 & \pu^2 \\ 
 \pd^2     & 0       & 0 \\
 \pu^2     & 0       & 0 \end{array} \right)\,,
\eeq
is singular. Nevertheless useful recursion relations 
can still be derived from eq.~(\ref{baseidentity}).
By choosing $y_i$ which satisfy the relation, $\sum_i S_{ji} y_i = a_j$
\beq
a=\left( 
\begin{array}{c}
 1 \\
 0 \\
 0 \end{array} \right),\;\;\;
y=\left( 
\begin{array}{c}
 0 \\
 \frac{\alpha}{\pd^2} \\
 \frac{1-\alpha}{\pu^2}
\end{array} \right)\,,
\eeq
and choosing the parameter $\alpha$ such that $\sum_i^3 y_i =0$,
we derive the recursion relation valid for $\nu_1>1$ and $p_1^2 \ne p_2^2$
\beq
\label{eq:recur2ma}
I (D;\nu_1,\nu_2,\nu_3) = \frac{1}{\pu^2-\pd^2} \frac{1}{(\nu_1-1)}
\Big[I (D-2;\nu_1-1,\nu_2-1,\nu_3)-I (D-2;\nu_1-1,\nu_2,\nu_3-1)\Big]\,.
\eeq
This relation lowers $D/2$ by one unit and $\sigma$ by two units and
is shown by a green (dashed) line in Fig.~\ref{fig:triangle2}.

By making the choice for $a$,
\beq
a=\left( 
\begin{array}{c}
 0 \\
 0 \\
 0 \end{array} \right),\;\;\;
y=\left( 
\begin{array}{c}
     0 \\
  \alpha \pu^2 \\
 -\alpha \pd^2
\end{array} \right)\,,
\eeq
we derive a further relation 
\beq
\label{eq:recur2mb}
I(D;\nu_1,\nu_2,\nu_3) = \frac{1}{D-1-\sigma} \frac{1}{\pd^2-\pu^2}
\Big[\pu^2\; I(D-2;\nu_1,\nu_2-1,\nu_3)-\pd^2\; 
I(D-2;\nu_1,\nu_2,\nu_3-1) \Big]\,,
\eeq
valid for $p_1^2 \ne p_2^2$.  This relation lowers $D/2$ by one and
$\sigma$ by one and is shown by a blue (solid) line in
Fig.~\ref{fig:triangle2}.

Note that there is a potentially dangerous case when $\nu_1=1$ and $D
= \sigma+1-2\e$, however if we restrict our attention to $n=0$ or
$n=-1$, this possibility can only occur for $D = 4-2\e$ and $\sigma=3$
which is a basis integral.

Finally we note that in the example considered later, Higgs decay to
four partons, $p_1^2 = p_2^2$ can never occur for two-mass triangles.
The degenerate case $p_1^2 = p_2^2$ can however also be treated easily
if needed.  For instance, by solving $S_{ij} y_j = a_j$ with $\sum_j
y_j =0$
\beq
a=\left( 
\begin{array}{c}
 0 \\
 0 \\
 0 \end{array} \right),\;\;\;
y=\left( 
\begin{array}{c}
     0 \\
  \alpha  \\
 -\alpha 
\end{array} \right)\,,
\eeq
one obtains 
\beq
I(D;\nu_1,\nu_2,\nu_3)=-I(D;\nu_1,\nu_2+1,\nu_3-1)\,. 
\eeq
This relation can be used to reduce the triangle to generalized
self-energies.

\subsection{The standard algorithm}
\label{sec:algo}
We summarize here the detailed steps of the algorithm as we
implemented it. In many cases there is freedom in the choice of the
particular recursion relation used.

The set of master integrals, which are the end-points of the
recursion relations, consists of 
\begin{itemize}
  \item all tadpoles $I(D;\nu_1)$ for arbitrary $D$ and
  $\nu_1$\,;
  \item all two-point functions $I(D;\nu_1,\nu_2)$ for arbitrary $D$ and
  $\{\nu_k\}$\,;
\item all one-mass triangles  $I(D;\nu_1,\nu_2,\nu_3)$ for arbitrary $D$ and
  $\{\nu_k\}$\,;
\item two-mass triangles  $I(D=4-2\e;1,1,1)$\,;
\item three-mass triangles  $I(D=4-2\e;1,1,1)$\,;
\item all boxes $I(D=4-2\e;1,1,1,1)$\,;
\item all pentagons $I(D=6-2\e;1,1,1,1,1)$\,.
\end{itemize}

For any given generalized scalar integral the algorithm proceeds as
follows:
\begin{itemize}
\item check if the integral has already been computed, if so read in
  the value of the integral, which has been previously stored;
\item check if the integral is a basis integral, if so evaluate its
  complex Laurent expansion using the analytic results (see Appendix~\ref{integrals}) and store the result; 
\item if $N=3$ and $\mbox{det}(S) = 0$ apply eq.~\eqref{eq:recur2mb}
  (reduction of two-mass triangles);
\item if $\sigma = N =5$ and $D=4-2\e$ apply eq.~\eqref{eq:recursion2a};
\item if $\sigma-N+2-\e=\frac{D}{2}$ apply eq.~\eqref{eq:recursion1};
\item if $\sigma-N+3-\e=\frac{D}{2}$ and $\sigma > N$ apply
  eq.~\eqref{eq:recursion3};
\item if $\sigma-N+3-\e=\frac{D}{2}$ and $\sigma = N$ apply
  eq.~\eqref{eq:recursion2};
\item if $\sigma-N+4-\e\le\frac{D}{2}$ 
apply eq.~\eqref{eq:recursion2}.
\end{itemize}

This procedure is repeated until all generalized scalar integrals have
been evaluated.

%------------------------------------------------------------------------
\section{The modified algorithm for exceptional kinematic regions}
\label{exceptional}
The numerical implementation of the recursion algorithm as outlined in
the previous section assumes that the kinematic matrix $S$ is
invertible for $N=4, 5$ and the quantity $B$ is not zero. If either
$B$ or $\det(S)$ is close to zero the numerical algorithm of section
III becomes unstable.

For these exceptional kinematic regions of phase space we need to
develop numerically sound procedures. Such exceptional regions are usually
associated with thresholds (i.e. a massive particle produced at rest)
or planar events (i.e. a linear dependence between the particle
vectors).

The exceptional kinematic regions are characterized by a vanishing
Gram determinant.  The Gram matrix is defined as (assuming $q_N=0$)
\beq
G_{ij}=2\, q_i\cdot q_j;\ \quad (i,j=1,2,\ldots, N-1)\ ,
\eeq
and is closely related to the kinematic matrix
\beq \label{GSconnection}
S_{ij}=(q_i-q_j)^2=\frac{1}{2}G_{ii}+\frac{1}{2}G_{jj}-G_{ij};\ 
\quad G_{iN}=G_{Ni}\equiv 0\ .
\eeq
The determinants of $S$ and $G$ are related through the $B$ quantity of
eq.~\eqref{eq:bidef}
\beq
\det(G) = (-1)^{N-1} B \det(S)\,.
\eeq
where $\det(G)$ denotes the determinant of the $(N-1)\times (N-1)$ submatrix.
We introduce some arbitrary small cutoffs $G_0, S_0$. 
We then distinguish 4 cases:
\begin{itemize}
\item Both $\det(S)>S_0$ and $\det(G)>G_0$:\newline In this case we
  apply the standard recursion relations of section III.
\item The $\det(S)\leq S_0$ and $\det(G)> G_0$ region:\newline This
  region is in fact not a kinematic exceptional region. We will
  rewrite the recursion relation in terms of the invertible $G$
  matrix. This gives an alternative set of recursion relations which
  do not depend on the kinematic matrix $S$.
\item The $\det(S)>S_0$ and $\det(G)\leq G_0$ region:\newline In this
  case the inverse of $S$ is still defined, while the $B$ parameter
  becomes small. This renders the recursion relations
  eqs.~\eqref{eq:recursion2} and~\eqref{eq:recursion3} unusable.  In
  ref.~\cite{Giele:2004ub} we already outlined a method to deal with
  this situation by rewriting the affected recursion relations as an
  expansion in the small parameter $B$.
\item The region where both $\det(S)\leq S_0$ and $\det(G)\leq
  G_0$:\newline In this region all recursion relations of section III
  have to be rewritten as expansions in the smallest eigenvalue of the
  kinematic matrix $S$.
\end{itemize}

\subsection{The region $\det(S)\leq S_0$ and $\det(G)> G_0$}
The numerical procedure of section III becomes unstable because the
inverse of the kinematic matrix $S$ is ill defined. We encountered
exactly this condition for the double off-shell triangles. As we saw
in that case two types of equations can be derived which we need to
generalize to boxes and pentagons.
 
The first set of equations is obtained by looking for the eigenvectors
associated with the smallest eigenvalues of $S$. If the determinant is
exactly zero one can find a non trivial eigenvector $y$ in the
null-space of $S_{ij}$ and use the base identity
eq.~\eqref{baseidentity} to derive
\beq
0=
-\sum_{i=1}^N y_i I(D-2;\{\nu_k-\delta_{ik}\}_{k=1}^N)
-\left(D-1-\sigma\right)\left(\sum_{i=1}^N y_i\right)
I(D;\{\nu_k\}_{k=1}^N)\,.
\eeq
If $\sum_{i=1}^N y_i \ne 0$ one obtains the relation 
\beq
\label{eq:recurdets1}
I(D;\{\nu_k\}_{k=1}^N) = -\frac{1}{D-1-\sigma}
\sum_{i=1}^N \frac{y_i}{\sum_{i=1}^N y_i} I(D-2;\{\nu_k-\delta_{ik}\}_{k=1}^N)\,, \eeq
which reduces both $D$ and $\sigma$ (and possibly $N$), while keeping
$n$ fixed.

If $\sum_{i=1}^N y_i = 0$ one obtains 
\beq
\label{eq:recurdets2}
I(D;\{\nu_k\}_{k=1}^N) =
-\sum_{i\ne l} \frac{y_i}{y_l} 
I(D;\{\nu_k-\delta_{ik} +\delta_{lk} \}_{k=1}^N)\,.
\eeq
This relation is not very efficient since it reduces neither $D$ nor
$\sigma$. However repeated use of eq.~(\ref{eq:recurdets2}) with $l$
fixed, is guaranteed to reduce $N$.

We now discuss how to extend eqs.~\eqref{eq:recurdets1}
and~\eqref{eq:recurdets2} to cases where the determinant is small, but
non zero.  In this case one computes the eigenvector $y$ of the matrix
$S_{ij}$ which has the smallest eigenvalue. The terms
\beq
\sum_{j=1}^N\left(\sum_{i=1}^N S_{ji}y_i\right)\nu_j 
I(D;\{\nu_k+\delta_{kj}\}_{k=1}^N)\,,
\eeq
in eq.~\eqref{baseidentity} give only a small contribution to the equation.
This means that, in the case $\sum_{i=1}^N y_i \ne 0$, one can rewrite
eq.~\eqref{baseidentity} as
\beqa \label{eq:detSexp1}
I(D;\{\nu_k\}_{k=1}^N) =& - &\frac{1}{D-1-\sigma}
\sum_{j=1}^N \frac{y_j}{\sum_{i=1}^N y_i} I(D-2;\{\nu_k-\delta_{kj}\}_{k=1}^N)
\nn
&-& \frac{1}{D-1-\sigma}
\sum_{j=1}^N\frac{\sum_{i=1}^N S_{ji}y_i}{\sum_{i=1}^N y_i}\nu_j 
I(D;\{\nu_k+\delta_{kj}\}_{k=1}^N)\,.
\eeqa
This equation allows one to reduce the integral on the l.h.s.  to a
sum of ``simpler integrals'' (the first term in the r.h.s.), i.\ e.\ 
integrals with lower $D$, lower $\sigma$ and possibly lower $N$, plus
a sum of more ``complicated integrals'' (the second term in the
r.h.s.), which have the same $D$ and higher $\sigma$, but which are
suppressed by the small coefficients $\sum_i S_{ji} y_i$.
Notice that since these integrals have exactly the same kinematic
matrix as the starting integral, the computation of these integrals
will make use of the same relation eq.~\eqref{eq:detSexp1}, with only
modified $D$ and exponents $\{\nu_k\}_{k=1}^N$, giving again simpler
integrals and ``more complicated integrals'' which are suppressed by
the square of the small coefficients $\sum_i S_{ji} y_i$. At some
point the small correction terms will be less than the desired
accuracy and the reduction can be terminated.

Similarly, in the case $\sum_{i=1}^N y_i$ close to zero, one obtains 
\beqa
\label{eq:detSexp2}
I(D;\{\nu_k\}_{k=1}^N) &=&  
-\sum_{i \ne l} \frac{y_i}{y_l} 
I(D;\{\nu_k-\delta_{ik} +\delta_{lk} \}_{k=1}^N) \nn
&-&\sum_{j=1}^N\frac{\sum_{i=1}^N 
S_{ji}y_i}{y_l}\left(\nu_j+\delta_{jl}\right) 
I(D+2;\{\nu_k+\delta_{kj}+\delta_{lk} \}_{k=1}^N)\nn
&-& \left(D-\sigma\right)\frac{\sum_{i=1}^N y_i}{y_l}
I(D+2;\{\nu_k +\delta_{lk}\}_{k=1}^N)
\,,
\eeqa
where now both the second and the third term in the r.h.s. are
small corrections to the result. For both integrals the kinematic
matrix is again the same as the starting one, so that a repeated
application of this same identity allows one to achieve the desired
accuracy in the answer.

Notice, that if $D=1+\sigma - 2\e$ in eq.~\eqref{eq:detSexp1}, then
the prefactor is of $\cO{1/\e}$. Therefore to have a correct result
for $I(D;\{\nu_k\}_{k=1}^N)$ up to $\cO{1}$ one needs the
$\e$-expansion of the integrals in the r.h.s. up to $\cO{\e}$.

To avoid $\cO{\e}$ expansions in the case $D=1+\sigma-2\e$ we derive
an alternative set of recursion equations.
Using eq.~\eqref{GSconnection} we write
\beq
\sum_{j=1}^N S_{ij} y_j=\frac{1}{2}G_{ii}\sum_{j=1}^N y_j
+\frac{1}{2}\sum_{j=1}^{N-1} G_{jj} y_j - \sum_{j=1}^{N-1} G_{ij} y_j\,,
\eeq
where we parametrized the scalar integral such that $q_N=0$, i.e.
$G_{iN}=0$. This means we are free to determine the values of $y_i$
($i=1,2,\ldots,N-1$) by solving $\sum_{j=1}^{N-1} G_{ij} y_j = r_i$
for any non-zero vector $r$ of our choosing and to define
$y_N=-\sum_{j=1}^{N-1} y_j$. We now have
\beq
\sum_{j=1}^N S_{ij} y_j=C(r)\times (1,1,\ldots,1)-r_i
\eeq
where $C(r)=\frac{1}{2}\sum_{j=1}^{N-1} G_{jj} y_j=\sum_{j=1}^{N-1}
q_j^2 y_j =\sum_{i,j=1}^{N-1} q_i^2 G_{ij}^{-1} r_j$

The new base equation now becomes (with $q_N=0$)
\beqa
\sum_{j=1}^{N-1} r_j\nu_j I(D;\{\nu_l+\delta_{lj}\}_{l=1}^N)&=&
\sum_{i,j=1}^{N-1} G_{ij}^{-1} r_j 
\left(I(D-2;\{\nu_l-\delta_{li}\}_{l=1}^N)
-I(D-2;\{\nu_l-\delta_{lN}\}_{l=1}^N)\right)\nn &&
-C(r) I(D-2;\{\nu_l\}_{l=1}^N)
\eeqa

To derive usable recursion relations we chose $r_i^{(k)}=\delta_{ik}$
($k=1,\dots,N-1$) and find the set of $(N-1)$ recursion relations
\beqa
\label{recurK}
(\nu_k-1) I(D;\{\nu_l\}_{l=1}^N)&=&
\sum_i^{N-1} G_{ik}^{-1} 
\left(I(D-2;\{\nu_l-\delta_{li}-\delta_{lk}\}_{l=1}^N)
-I(D-2;\{\nu_l-\delta_{lN}-\delta_{lk}\}_{l=1}^N)\right)\nn &&
-C_k I(D-2;\{\nu_l-\delta_{lk}\}_{l=1}^N)
\eeqa
with $C_k=\sum_{i=1}^{N-1} q_i^2 G_{ik}^{-1}$.

By summing over $k$ we derive the recursion relation for the case that
$\nu_N\neq 1$ 
\beqa
\label{recurN}
(\nu_N-1) I(D;\{\nu_l\}_{l=1}^N)&=&
-\sum_i^{N-1} g_i
\left(I(D-2;\{\nu_l-\delta_{li}-\delta_{lN}\}_{l=1}^N)
-I(D-2;\{\nu_l-2\delta_{lN}\}_{l=1}^N)\right)\nn &&
+(C-1) I(D-2;\{\nu_l-\delta_{lN}\}_{l=1}^N)
\eeqa
with $C=\sum_{k=1}^{N-1} C_k$ and $g_i=\sum_{k=1}^{N-1} G_{ik}^{-1}$.

Note that these equations can be applied only if at least one of the
$\nu_i>1$.
The cases where all $\nu_i=1$ and $D=1+\sigma-2\e$ correspond always
to basis integrals, so that all cases are covered by the recursion
relations presented here.~\footnote{In our current algorithm we chose
  to apply eq.~\eqref{eq:detSexp1} for triangles even if
  $D=1+\sigma-2\e$. In this case we need the expansion of two-point
  functions up to $\cO{\e}$.  Suitable results are reported in
  Appendix~\ref{integrals}.}

\subsection{The region $\det(S)>S_0$ and $\det(G)\leq G_0$}
In the case where $B$ vanishes but det$(S_{ij}) \ne 0$ problems arise
only if one is to use relations eqs.~\eqref{eq:recursion2}
or~\eqref{eq:recursion3}, while eqs.~\eqref{eq:recursion1}
or~\eqref{eq:recursion2a} can be used in a straightforward way.

For small $B$ one can simply use eq.~\eqref{eq:recursion2a}
\begin{equation}
\label{eq:Bexp}
I(D;\{\nu_k\}_{k=1}^N)=
\sum_{i=1}^Nb_iI(D;\{\nu_k-\delta_{ik}\}_{k=1}^N)
+\left(D+1-\sigma\right)\,B\,I(D+2;\{\nu_k\}_{k=1}^N)\,,
\end{equation}
where the first term gives rise to simpler integrals (lower $\sigma$
and possibly lower $N$), while the second term gives a small
correction to the result and is computed by applying this same
recursion relation iteratively until the desired accuracy is reached.

\subsection{The region $\det(S)\leq S_0$ and $\det(G)\leq G_0$}
First we consider phase space points where the Gram determinant
exactly vanishes. In this case we have a non-trivial solution of the
equation
\beq
\label{eq:zvec}
\sum_{j=1}^{N-1} G_{ij} z_j=0\,.
\eeq
Setting $z_N=-\sum_{i=1}^{N-1} z_i$ leads to
\beq
\label{Cdef}
\sum_{j=1}^N S_{ij} z_j=C\times\left(1,1,\ldots,1\right)\,,\qquad 
C\equiv \frac12 \sum_{k=1}^{N-1} G_{kk} z_k = \sum_{k=1}^{N-1} q_k^2 z_k\,.
\eeq

If $C>C_0$ we can rescale $z_i$: $r_i=z_i/C$.  From
eqs.~\eqref{baseidentity} and~\eqref{Cdef}) we obtain
\begin{equation}
\label{eq:recurmod3}
I(D;\{\nu_k\}_{k=1}^N) = 
\sum_{i=1}^N r_i I(D;\{\nu_k-\delta_{ik}\}_{k=1}^N)\,.
\end{equation}

If $\det(G_{ij})$ is close to zero, the equation $S_{ij} r_j =1$ will
be approximate, leaving small correction terms $\Delta_i$
\begin{equation}
\label{eq:rvecapp}
S_{ij} r_j =\left(1,1,\ldots,1\right) + \Delta_i.
\end{equation}
By explicit construction it is possible to keep exact the condition
$\sum_{j=1}^N r_j =0$.  In this case the expanded recursion relation
becomes
\beqa
\label{eq:recurmod3exp}
I(D;\{\nu_k\}_{k=1}^N) &= &\sum_{i=1}^N r_i 
I(D;\{\nu_k-\delta_{ik}\}_{k=1}^N)\nn
&+&\sum_{i=1}^N \Delta_i \nu_i 
I(D+2;\{\nu_k+\delta_{ik}\}_{k=1}^N)\>. 
\eeqa

If $C\leq C_0$ we cannot rescale the vector $z_i$. Instead we write an
expansion
\beqa
\label{eq:recurmod4exp}
I(D;\{\nu_k\}_{k=1}^N) &=&
-\sum_{i\ne l} \frac{z_i}{z_l} I(D;\{\nu_k-\delta_{ik} +\delta_{lk} \}_{k=1}^N)\nn &&
-\sum_{j=1}^N \frac{\sum_{i=1}^N S_{ji}z_i}{z_l} (\nu_j+\delta_{jl})
I(D+2;\{\nu_k+\delta_{kj}+\delta_{lk}\}_{k=1}^N)\,.
\eeqa
We finally note that the expanded recursion relations change the value
of $n$ defined in eq.~\eqref{eq:ndef}. Since more steps might be
needed to reach the desired accuracy these relations can lead outside the
normal IR and UV boundaries, giving rise to deep-IR or deep-UV integrals.

%----------------------------------------------------------------------
\subsection{Summary of the modified algorithm}
\label{sec:modalgo}
The set of standard and expanded recursion relations is over-complete.
Therefore there is some freedom in most steps of the algorithm.  We
summarize here the detailed steps of the modified algorithm, as we
implemented it. While the procedure works, it is yet to be optimized.

The set of master integrals is the same as the one listed in
section~\ref{sec:algo}. Notice that, as explained above, two-point
functions need to be expanded up to $\cO{\e}$ and basis integrals out
of the standard IR/UV boundaries will appear.

We rescale the hard event by a typical hard scale of the process. In
order to measure the closeness to the exceptional momentum
configuration we introduce a small parameter $\zeta \ll 1$ (which we
fixed at $\zeta=10^{-6}$ in the numerical examples).

Given any scalar integral, the modified algorithm proceeds as follows:
\begin{itemize}
\item check if the integral has already been computed; if so, read in
  the value of the integral, which has been previously stored;
\item check if the integral is a basis integral; if so, evaluate its
  complex Laurent expansion and store the result;
\item if all eigenvalues of $S_{ij}$ are larger than ${\zeta} $ the
  standard algorithm of section~\ref{sec:algo} can be applied, with
  the following modifications:
  \begin{itemize}
  \item[{$\cdot $}] if $\sigma = N =5$ and $D=4-2\e$, or
    $\sigma-N+2-\e> \frac{D}{2} $ apply eq.~\eqref{eq:recursion2a};
  \item[{$\cdot $}] if $\sigma-N+2-\e=\frac{D}{2}$ apply
    eq.~\eqref{eq:recursion1};
  \item[{$\cdot $}] if $\sigma-N+3-\e=\frac{D}{2}$ and $B > \zeta$
    apply eq.~\eqref{eq:recursion3};
  \item[{$\cdot $}] if ($\sigma-N+4-\e \le \frac{D}{2}$ or 
($\sigma-N+3-\e=\frac{D}{2}$ and $\sigma = N$))
and $B > \zeta$
    apply eq.~\eqref{eq:recursion2};
  \item[{$\cdot $}] if  
$\sigma-N+3-\e=\frac{D}{2}$ and $\sigma = N$
and $B > \zeta$
    apply eq.~\eqref{eq:recursion2};
  \item[{$\cdot $}] if $\sigma-N+4-\e \le \frac{D}{2}$ 
and $B > \zeta$
    apply eq.~\eqref{eq:recursion2};
  \item[{$\cdot $}] if $\sigma-N+3-\e\le \frac{D}{2}$ and $B < \zeta$
    apply eq.~\eqref{eq:recursion2a}.
  \end{itemize}
  
\item if some eigenvalue of $S_{ij}$ is smaller than $\zeta$, but all
  eigenvalues of $G_{ij}$ are larger, the event is not really
  exceptional, however standard relations of sec.~\ref{sec:algo} can
  not be applied since the kinematic matrix has no inverse,
  accordingly
  \begin{itemize}
  \item[{$\cdot $}] if $D\ne \sigma+1-2\e$ or $N=3$ apply
    eq.~\eqref{eq:detSexp1};
  \item[{$\cdot $}] if $D= \sigma+1-2\e$, $N\ne 3$ and some $\nu_k \ne
    1 $ ($k=1,\dots,N-1$) apply eq.~\eqref{recurK};
  \item[{$\cdot $}] if $D= \sigma+1-2\e$, $N \ne 3$ and $\nu_N \ne 1$
    apply eq.~\eqref{recurN}.
  \end{itemize}
  
\item if both $S_{ij}$ and $G_{ij}$ have eigenvalues smaller than
  $\zeta$: the event is considered exceptional. In this case
  \begin{itemize}
  \item[{$\cdot $}] if $C > \sqrt{\zeta}$ apply
    eq.~\eqref{eq:recurmod3exp} (where $C$ is defined in
    eq.~\eqref{Cdef});
  \item[{$\cdot $}] if $C \le \sqrt{\zeta}$ apply
    eq.~\eqref{eq:recurmod4exp}.
  \end{itemize}
\end{itemize}
The number of terms included in the expanded relations is limited by
the requirement that $k_i^{\rm niter-1} > \zeta$. Here $n_{\rm iter}$
denotes the number of iterations and $k_i$ are the small coefficients
in
eqs.~\eqref{eq:recursion2a},~\eqref{eq:detSexp1},~\eqref{eq:recurmod3exp}
and ~\eqref{eq:recurmod4exp}, which parameterize the departure from the
exceptional phase space points.  Concretely, the $k_i$ are the
prefactors of the integrals $I$ in the second term in the r.h.s of
those equations.

%--------------------------------------------------------------------
\subsection{Some explicit results for $H\to q \bar q q^\prime \bar{q}^\prime$}
An ideal process to illustrate expansions close to exceptional
momentum configurations is the decay 
\beq
\label{eq:Hqarbv}
H (-p_5)\to q(p_1)\> \bar q(p_2)\> q^\prime(p_3)\> \bar{q}^\prime(p_4)\,.
\eeq  
This is a non-trivial example, with $N=5$ and one massive external
momentum, where the analytical result has recently become
available~\cite{Ellis:2005qe}.  (Alternatively one could study the
process $H \to q\bar q q \bar q$ which is also given in
ref.~\cite{Ellis:2005qe}).  A close inspection of this result reveals
immediately that it is free of any spurious singularities and
therefore its evaluation is stable for any phase space
point.\footnote{Remember we are interested in separate jet-production,
  i.\ e.\ we always assume $|s_{ij}| > s_{\rm min} >0$.}

We present here a comparison between the numerical and the analytical
result for phase space points approaching exceptional momentum
configurations.

For the Higgs decay process we are considering, the determinant of the
kinematic matrix defined in eq.~(\ref{Sdef}) is given by
\beq
\label{eq:detS}
\det(S_{ij}) = 2 s_{12} s_{23} s_{34}(s_{123} s_{234}- p_5^2 s_{23})\,,
\eeq 
where $s_{ijk}\equiv s_{ij}+s_{jk}+s_{ik}$ and the momentum
assignments are given in eq.~\eqref{eq:Hqarbv}.
It is straightforward to see that if $p_4$ is a linear combination of
$p_1$ and $p_2+p_3$, the determinant of the kinematic matrix vanishes,
\beq 
p_4 = \alpha p_1 +\beta p_2 + \gamma p_3 \quad\mbox{with}\quad \alpha=
-\frac{\beta \gamma s_{23}}{\beta s_{12} +\gamma s_{13}} \quad\mbox{and}\quad
\gamma = \beta \quad \Rightarrow \det(S_{ij}) = 0\,,
\eeq
where the condition on $\alpha$ ensures that $p_4$ is massless.  It is
then possible to approach the limit $\det(S_{ij}) = 0$ slowly, by
choosing $\gamma = \beta +\delta$, where $\delta$ is a small variation
such that $\delta \ll \alpha, \beta,\gamma$.

As an illustration we present results for the following
exceptional momentum configurations, $(E,p_x,p_y,p_z)$,
\beq
\label{eq:momentumchoice}
\begin{array}{lllll}
p_1=& (-0.50000000000,0.00000000000,0.00000000000,-0.50000000000)\,,\\
p_2=& (-0.50000000000,0.00000000000,0.00000000000,+0.50000000000)\,,\\
p_3=& (+0.10000000000,0.05000000000,0.00000000000,+0.08660254037)\,,\\ 
p_4=& \alpha p_1 +\beta p_2 + \gamma p_3\qquad \alpha=
-\beta \gamma s_{23}/(\beta s_{12} +\gamma s_{13})\qquad \beta = 1/3 \qquad\gamma = \beta +\delta\,,\\
p_5=&-p_1-p_2-p_3-p_4\,,
\end{array}
\eeq 
where $\beta$ is an arbitrary number $\cO{1}$ and we vary $\delta$ in
the range $-50<\ln\delta< 0$.

\begin{figure}
\begin{minipage}[t]{0.32\textwidth}
  \includegraphics[width=1.13\textwidth]{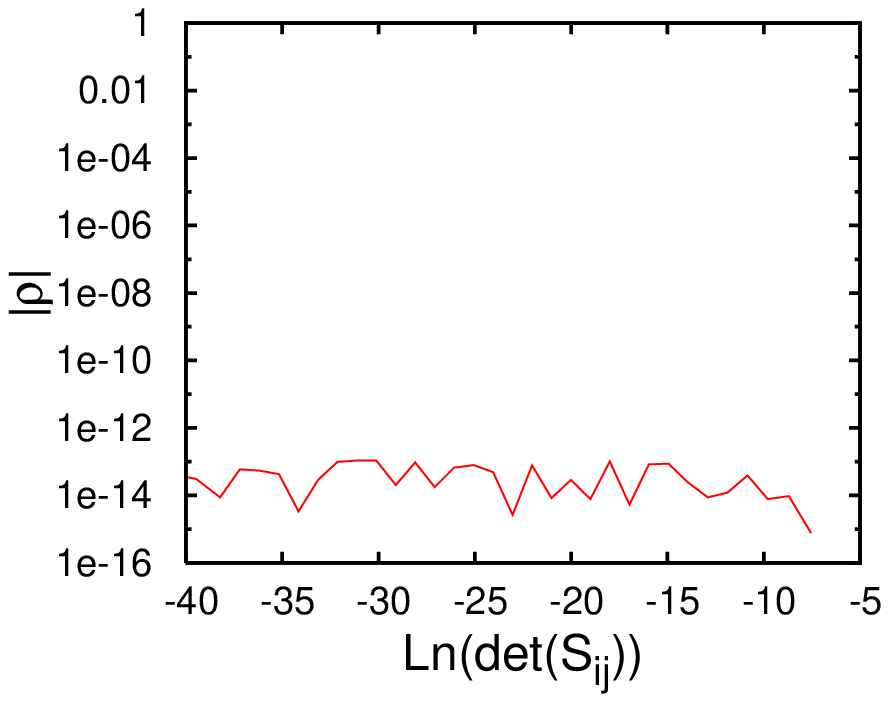}
\end{minipage}
\begin{minipage}[t]{0.32\textwidth}
  \includegraphics[width=1.13\textwidth]{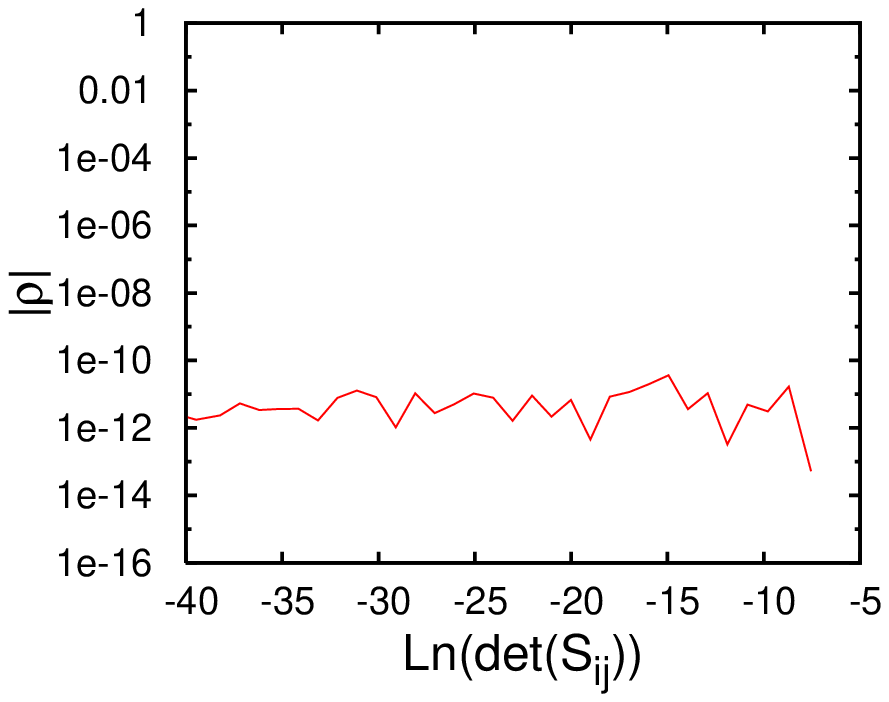}
\end{minipage}
\begin{minipage}[t]{0.32\textwidth}
  \includegraphics[width=1.13\textwidth]{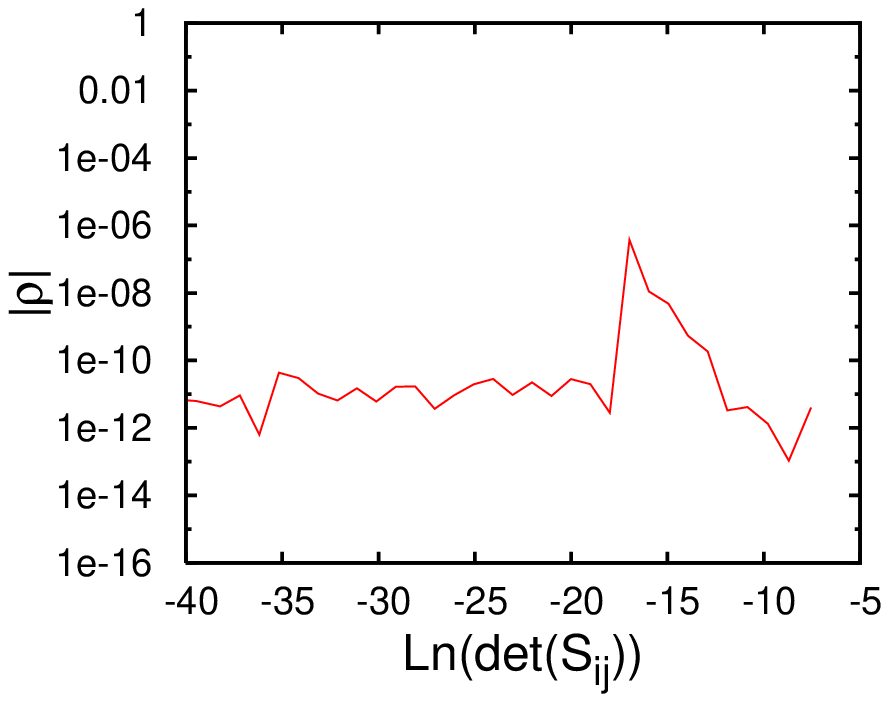}
\end{minipage} 
\caption{\label{fig:detS} Relative accuracy $|\rho|$ for the $1/\epsilon^2$ pole (left), the $1/\epsilon$ pole (center) and the constant part (right) of the one-loop amplitude squared for $H\to q \bar{q} q' \bar{q}'$ as a function of the determinant of the kinematics matrix $S_{ij}$.}
\end{figure}

In Fig.~\ref{fig:detS} we plot the absolute value of the relative
accuracy
\begin{equation}
\rho  = \frac{A_{V,N}-A_{V,A}}{A_{V,A}}\,,
\end{equation}
where, as defined in~\cite{Ellis:2005qe}, $A_{V,N}$ and $A_{V,A}$ denote
the numerical and analytical result for the full one-loop amplitude
squared for the process $H\to q\bar q q' \bar q'$.

The matrix $S_{ij}$ denotes here the kinematic matrix for the standard
ordering of momenta, as defined in eq.~\eqref{Sdef}. In the evaluation
of a full one-loop amplitude one encounters $(N-1)!$ such matrices,
which are obtained by permuting $(N-1)$ momenta keeping one arbitrary
momentum fixed
\beq \label{SdefP}
S_{ij}^\sigma=\left(q_{\sigma(i)}-q_{\sigma(j)}\right)^2,
\eeq
where $\sigma(i)$ denotes the permutation of index $i$.  Additionally,
external momenta are contracted together, giving reduced matrices.
Whenever any such matrix is close to singular, the modified recursion
relations are used as explained in sec.~\ref{sec:modalgo}. Therefore
plotting the accuracy as a function of just one determinant is only
indicative.

From Fig.~\ref{fig:detS} we see that the numerical answer is
well-behaved when approaching exceptional phase space points,
$\det(S_{ij})\to0$. Since the closeness to the exceptional phase space
point is the expansion parameter in the modified recursion relations,
an extremely good accuracy can be reached very close to singular
points by including only a few terms in the expansion. In the
intermediate region, where the expansion parameter is larger, one has
to include more terms in the expansion. This however means computing a
larger number of integrals with higher $D$ and $\sigma$.
In the current example, eq.~\eqref{eq:Hqarbv} a relative accuracy of
$10^{-6}$ was achieved, limited by the computer memory. A greater
accuracy could be obtained by allocating more computer memory.

We now study the case $B\to0$. If the four-vector $p_4$ is a linear
combination of $p_1, p_2$ and $p_3$ then for any choice of
$\alpha,\beta$ and $\gamma$ which keeps $p_4^2=0$, we find $B=0$.
However for the process in eq.~\eqref{eq:Hqarbv} the only pentagon
integrals appearing in the numerical answer are scalar integrals in
$D=4-2\epsilon$ with $\sigma =5$. These integrals can be reduced using
the standard eq.~\eqref{eq:recursion2a} which holds regardless of
whether $B=0$ or not.

To study the $B$ expansion we therefore choose specific phase space
points close to $\gamma=-1$. In this case the coefficient $B$ of the
reduced $4\times 4$ kinematic matrix, which is obtained by setting
$\nu_3 = 0$, vanishes and an expansion in $B$ is needed.  We choose
the same momenta as in eq.~\eqref{eq:momentumchoice}, but fix
$\gamma=-1+\delta$ and vary $\delta$ in the range $-8<\ln\delta< 0$.

\begin{figure}
\begin{minipage}[t]{0.32\textwidth}
  \includegraphics[width=1.13\textwidth]{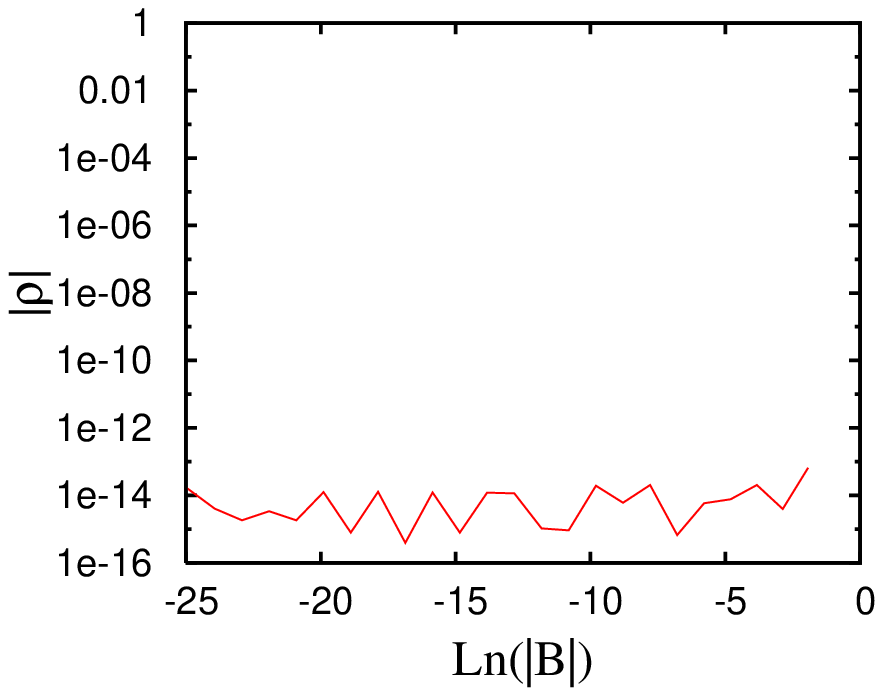}
\end{minipage}
\begin{minipage}[t]{0.32\textwidth}
  \includegraphics[width=1.13\textwidth]{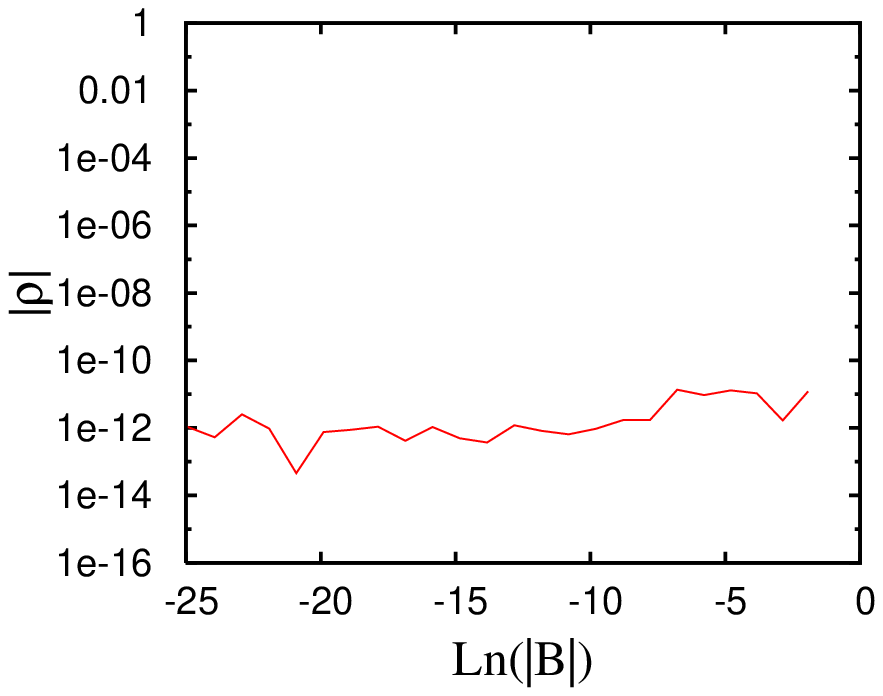}
\end{minipage}
\begin{minipage}[t]{0.32\textwidth}
  \includegraphics[width=1.13\textwidth]{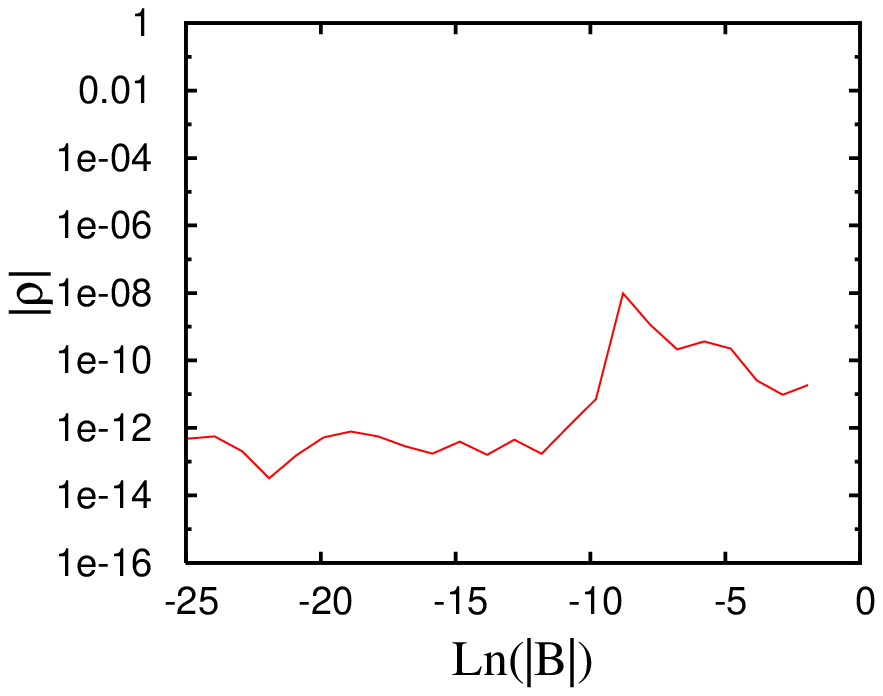}
\end{minipage} 
\caption{\label{fig:B} Relative accuracy $|\rho|$ for the $1/\epsilon^2$ 
  pole (left), the $1/\epsilon$ pole (center) and the constant part
  (right) of the one-loop amplitude squared for $H\to q \bar{q} q'
  \bar{q}'$ as a function of $\ln(|B|)$ for the box integral obtained
  by setting $\nu_3=0$.}
\end{figure}

Fig.~\ref{fig:B} shows then the absolute value of the relative
accuracy for the numerical one-loop amplitude squared for $H\to q\bar
q q' \bar q'$, plotted as a function of $\ln(B)$ of the reduced matrix
obtained by contracting the momenta $p_3$ and $p_4$ ($\nu_3 =0$).
Again, numerical results are stable close to exceptional momentum
configurations and a relative accuracy of $10^{-6}$ can be achieved.

%-----------------------------------------------------------------
\section{Outlook}
We presented here a semi-numerical method to calculate one-loop
corrections to scattering amplitudes with up to five external legs and
massless internal lines.
A variety of approaches to deal with the exceptional kinematic regions
have been developed. Further developments are possible, guided by
practical applications. We demonstrated that numerical methods can be
extended to kinematic regions such as threshold and planar events.

With these results we can integrate the virtual matrix element
of~\cite{Ellis:2005qe} over all of phase space to obtain the fully
differential Higgs boson plus two jet cross section at next-to-leading
order.

It is now straightforward to extend the calculation to other, yet
uncalculated, cross sections.  Examples of processes of current
experimental interest are di-boson plus one jet ($V_1,V_2,j$),
tri-boson production ($V_1,V_2,V_3$) and vector boson plus heavy quark
pairs ($V Q \bar{Q}$). The extension to the case with internal masses can be
implemented without much difficulty.

Furthermore, the recursive techniques are not restricted to 5 external
particles. An extension of the method to 6 or more external particles
is only limited by computer resources. To what extent this will be
feasible in practical applications still needs to be investigated.

Finally, because the entire calculation is done numerically we have
great flexibility in recalculating the cross sections in any
``scheme'' necessary.  For example, this flexibility should be helpful
in combining NLO calculations with shower Monte Carlo programs.

\appendix

\section{Integrals}
\label{integrals}

We report here expressions for all of the basis integrals needed as
end points for the recursion relations used in our approach. All of
these integrals can be found in the literature, but we find it useful
to collect the results here.

We first introduce some notations.  We will denote $I_N(D;\nu_1\dots
\nu_N)$ the master integral $I(D;\nu_1\dots \nu_N)$ with $N$ external
legs. The quantity $\cG$ which is equal to unity in 4 dimensions is
common to all integrals and is given by
\beq
\cG=\frac{\Gamma^2(1-\e)\Gamma(1+\e)}{\Gamma(1-2\e)} = 
\frac{1}{\Gamma(1-\e)} +O(\e^3)\,.
\eeq
We will frequently need the following expansions in $\e$
\beqa
\frac{\Gamma(n-\e)}{\Gamma(1-\e)}&=&\Gamma(n)
\left(1-\e H_1(n)+\frac{\e^2}{2}\left\{H_1^2(n)-H_2(n)\right\}\right)\,,
\nonumber\\
\frac{\Gamma(-n-\e)}{\Gamma(1-\e)}&=&\frac{(-1)^{n+1}}{\e\Gamma(n+1)}
\left(1-\e H_1(n+1)-\frac{\e^2}{2}\left\{H_1^2(n+1)+H_2(n+1)\right\}\right)\,,
\nonumber\\
\eeqa
where the generalized harmonic sum is defined by
\beq
H_k(n) \equiv \sum_{m=1}^{n-1}\frac{1}{m^k}\,. 
\eeq
The dilogarithm is defined as usual as
\beq
\li(x)\equiv - \int_{0}^{x} \frac{dt}{t} \; \ln(1-t) \, . 
\eeq
\subsection{Tadpole diagrams:  $I_1(D=2(\sigma+n)-2\e;\nu_1)$}
Within dimensional regularization tadpole integrals are zero
\beq
I_1(D;\nu_1)=\int\frac{d^D\ell}{i\pi^{D/2}}\frac{1}{d_1^{\nu_1}}=0\,.
\eeq
\subsection{Self energy diagrams: $I_2(D=2(\sigma+n)-2\e;\nu_1,\nu_2\,|\,\{\pu^2\})$}
\begin{figure}[t]
\includegraphics[angle=270,scale=0.65]{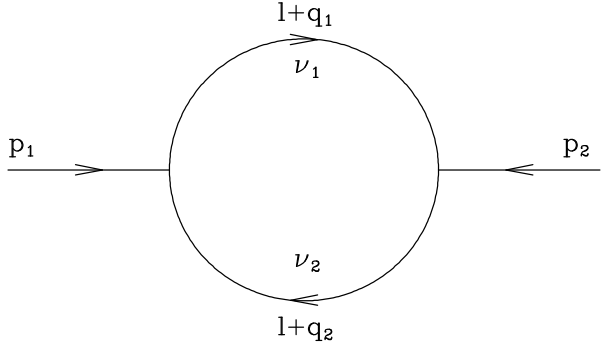}
\caption{The generic self energy diagrams.} 
\label{fig:sediag}
\end{figure}
All generalized scalar two-point integrals, shown in fig.~\ref{fig:sediag}, 
are master integrals. 
\beqa
\lefteqn{I_2(D=2(\sigma+n-\e);\nu_1,\nu_2\,|\,\{\pu^2\})=
\int\frac{d^D\ell}{i\pi^{D/2}}\frac{1}{d_1^{\nu_1}d_2^{\nu_2}}} \nn &=&
(-1)^{\sigma}(-\pu^2)^{n-\e}\frac{\Gamma(\nu_1+n-\e)\Gamma(\nu_2+n-\e)\Gamma(-n+\e)}
{\Gamma(\nu_1)\Gamma(\nu_2)\Gamma(\sigma+2n-2\e)}\,.
\eeqa
Depending on the value of $n$, defined in eq.~\eqref{eq:ndef}, we have different expansions.
The cases $n > 0$ and $n<-1$ are needed to deal with the exceptional momentum configurations.
\begin{itemize}
\item The UV case $n\geq 0$:
\beqa
\lefteqn{I_2(D=2(\sigma+n-\e);\nu_1,\nu_2\,|\,\{\pu^2\})
= \frac{(-1)^{\sigma+n}c_{\Gamma}\; \Gamma(\nu_1+n)\Gamma(\nu_2+n) \left(-\pu^2\right)^n}
{\Gamma(\nu_1)\Gamma(\nu_2)\Gamma(n+1) \Gamma(\sigma+2n)}
}\nn 
 &\times \Bigg\{&
\frac{1}{\e}
 +2 H_1(\sigma+2 n)+H_1(n+1)-H_1(\nu_1+n)-H_1(\nu_2+n)-\ln(-\pu^2)
\nonumber \\
&+& \frac{\e}{2} \Bigg[
\bigg(2 H_1(\sigma+2 n)+H_1(n+1)-H_1(n+\nu_1)-H_1(n+\nu_2)-\ln(-\pu^2)\bigg)^2
\nonumber \\
&& \phantom{\frac{\e}{2} }+ 4 H_2(\sigma+2n)+ H_2(n+1)-H_2(n+\nu_1)-H_2(n+\nu_2)
\Bigg] \Bigg\}+\cO{\e^2}\,.
\nonumber \\
\eeqa
\relax
\relax
\relax
\relax
\item The IR cases $n < 0$ 
\begin{enumerate}
\item $\nu_1 >  -n $, $\nu_2 >  -n $:
\beqa
\lefteqn{I_2(D=2(\sigma+n-\e);\nu_1,\nu_2\,|\,\{\pu^2\})
= \frac{(-1)^{\sigma}c_{\Gamma}\; 
\Gamma(-n) \Gamma(\nu_1+n)\Gamma(\nu_2+n) \left(-\pu^2\right)^n }
{\Gamma(\nu_1)\Gamma(\nu_2)\Gamma(\sigma+2n)}}
\nonumber \\
&\times&\Bigg\{1 + \e
\Big[2 H_1(\sigma + 2n)+
 H_1(-n) - H_1(\nu_1+n) - H_1(\nu_2+n) -\ln(-\pu^2)
\Big]\Bigg\}\nn
&+&\cO{\e^2}\,.
\eeqa
\relax
\relax
\relax
\relax
\relax
\item $\nu_1 >  -n $, $\nu_2 \le  -n $, $\nu_1 + \nu_2 >  -2 n $ :
\beqa
\lefteqn{I_2(D=2(\sigma+n-\e);\nu_1,\nu_2\,|\,\{\pu^2\})
= \frac{(-1)^{\nu_1+n+1}c_{\Gamma}\; \Gamma(\nu_1+n)\Gamma(-n) \left(-\pu^2\right)^n}
{\Gamma(\nu_1)\Gamma(\nu_2)\Gamma(1-n-\nu_2) \Gamma(\sigma+2n)}
}\nn 
&& \times
\Bigg\{\frac{1}{\e}
     +2 H_1(\sigma+2 n)+H_1(-n)-H_1(\nu_1+n)-H_1(1-\nu_2-n)-\ln(-\pu^2) \nn
&&+ \frac{\e}{2}
\Bigg[\bigg(2 H_1(\sigma+2 n)+H_1(-n)-H_1(\nu_1+n)-H_1(1-\nu_2-n)-\ln(-\pu^2) \bigg)^2
 \nn
&& \phantom{\frac{\e}{2} } + 4H_2(\sigma+2n)-H_2(-n)-H_2(\nu_1+n)+H_2(1-\nu_2-n)\Bigg]
\Bigg\}\nn
&+&\cO{\e^2}\,.
\eeqa
\relax
\relax
\relax
\relax
\relax
\relax
\relax
\item $\nu_1 >  -n $, $\nu_2 \le  -n $, $\nu_1 + \nu_2 \le  -2 n $ :
\beqa
&&I_2(D=2(\sigma+n-\e);\nu_1,\nu_2\,|\,\{\pu^2\})
= \frac{2 (-1)^{\nu_2+n }c_{\Gamma}\; \Gamma(\nu_1+n)\Gamma(-n)
\Gamma(1-\sigma-2n) (-\pu^{2})^n }
{\Gamma(\nu_1)\Gamma(\nu_2)\Gamma(1-n-\nu_2)}
\nn 
&&\times\Bigg\{ 1+ 
\e\Big[ 2H_1(1-\sigma-2 n) +H_1(-n)-H_1(\nu_1+n)-H_1(1-\nu_2-n) 
-\ln(-p^2) \Big]\Bigg\}\nn
&&+\cO{\e^2}\,.\nn
\eeqa
\relax
\relax
\relax
\relax
\relax
\relax
\relax
\relax
\item $\nu_1 \le  -n $, $\nu_2 \le  -n $ :
\beqa
&&I_2(D=2(\sigma+n-\e);\nu_1,\nu_2\,|\,\{\pu^2\})
= \frac{2 (-1)^{\sigma+1}c_{\Gamma}\; \Gamma(1-\sigma-2n)\Gamma(-n)}
{\Gamma(\nu_1)\Gamma(\nu_2)\Gamma(1-n-\nu_1) \Gamma(1-n-\nu_2)}
\left(-\pu^2\right)^n\nn 
&& 
\times\Bigg\{
\frac{1}{\e}
     +2 H_1(1-\sigma-2 n)+H_1(-n)-H_1(1-n-\nu_1)-H_1(1-\nu_2-n)-\ln(-\pu^2) \nn
&&+\frac{\e}{2}
\Big[(2 H_1(1-2 n-\sigma)+H_1(-n)-H_1(1-\nu_1-n)-H_1(1-\nu_2-n)-\ln(-\pu^2))^2
\nn
&-&4 H_2(1-2n-\sigma)-H_2(-n)+H_2(1-n-\nu_1)+H_2(1-n-\nu_2)
\Big]\Bigg\}\nn
&&+\cO{\e^2}\,.
\eeqa
\end{enumerate}
\end{itemize}

\subsection{Triangle diagrams: 
$I_3(D=2(\sigma+n-\e);\nu_1,\nu_2,\nu_3\,|\,\{\pu^2,\pd^2,\pt^2\})$}
\label{sec:Triangles}
\begin{figure}[t]
\includegraphics[angle=270,scale=0.75]{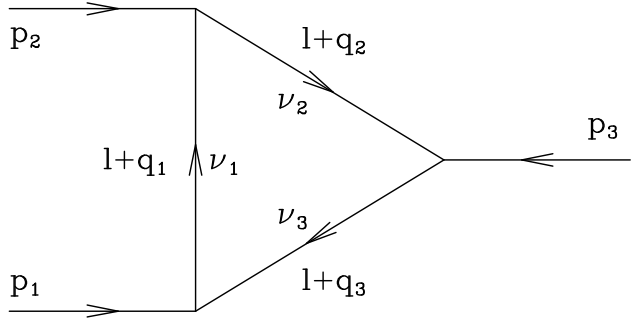}
\caption{The generic triangle diagram.}
\label{fig:trianglediag}
\end{figure}

In the case of the 3-point general scalar integrals, shown in
fig.~\ref{fig:trianglediag}, we distinguish three cases
depending on the number of off-shell legs.

\subsubsection{One off-shell leg}
The first case is the three-point generalized scalar integral with one
off-shell leg.  This is again a master integral.

The general expression for the  triangle with one off-shell leg
($\pu^2=\pd^2=0$) is given by~\cite{Anastasiou:1999ui}
\beqa \label{basictriangle}
&&I_3(D=2(\sigma+n)-2\e;\nu_1,\nu_2,\nu_3 |\{0,0,\pt^2\})
\nonumber \\
&=&(-1)^\sigma (-\pt^2)^{n-\e}
\frac{\Gamma(\nu_2+n-\e)}{\Gamma(\nu_2)}
\frac{\Gamma(\nu_3+n-\e)}{\Gamma(\nu_3)}
\frac{\Gamma(-n+\e)}{\Gamma(\sigma+2n-2\e)} 
\nonumber \\
&=&\cG\frac{(-1)^\sigma (-\pt^2)^{n-\e}}{(\sigma+2n-1)!}
\left(\frac{\Gamma(-n+\e)}{\Gamma(1+\e)}\right) 
\nonumber \\ &&\times
\left(\frac{\Gamma(\nu_2+n-\e)}{\Gamma(\nu_2)\Gamma(1-\e)}\right) 
\left(\frac{\Gamma(\nu_3+n-\e)}{\Gamma(\nu_3)\Gamma(1-\e)}\right) 
\left(\frac{\Gamma(1-2\e)\Gamma(\sigma+2n)}{\Gamma(\sigma+2n-2\e)}\right)\,.
\nonumber \\
\eeqa
Depending on the value of $n$, the integral is IR ($n=-1$), 
UV ($n=0$) or deep UV ($n\geq 1$) 
divergent. We will look at these three specific cases:
\begin{itemize}
\item {\bf The IR case}: $n=-1$ \newline
The generic integral now becomes
\beqa
\lefteqn{I_3^{IR}(D=2(\sigma-1-\e); \nu_1,\nu_2,\nu_3)
=-\cG\frac{(-1)^\sigma}{(\sigma-3)!}\frac{(-\pt^2)^{-\e}}{\pt^2}}
\nonumber\\
&\times&\left(\frac{\Gamma(\nu_2-1-\e)}{\Gamma(\nu_2)\Gamma(1-\e)}\right) 
\left(\frac{\Gamma(\nu_3-1-\e)}{\Gamma(\nu_3)\Gamma(1-\e)}\right) 
\left(\frac{\Gamma(1-2\e)\Gamma(\sigma-2)}{\Gamma(\sigma-2-2\e)}\right)\,.  
\eeqa
Depending on the values of $\nu_2$ and $\nu_3$ we can distinguish three cases
\begin{enumerate}
\item $\nu_2=\nu_3=1$:\ $\sigma=\nu_1+2$ \newline
This is the most singular case (${\cal O}(1/\e^2)$)
\beqa
&& I_{1m}^{IR}(D=2(\nu_1+1-\e); \nu_1,1,1)
= -\cG\frac{(-1)^{\nu_1}}{(\nu_1-1)!}\frac{1}{\pt^2}
\nonumber \\ 
&\times& \Bigg[
\frac{1}{\e^2}+\frac{1}{\e}\left(2H_1(\nu_1)-\ln(-\pt^2)\right)
\nonumber \\ 
& + &\frac{1}{2}\ln^2(-\pt^2)-2H_1(\nu_1)\ln(-\pt^2)
 +2\left(H_1^2(\nu_1)+H_2(\nu_1)\right)
\Bigg]+{\cal O}(\e)\,. \nonumber \\
\eeqa
\item ($\nu_2=\nu >1;\ \nu_3=1$) or ($\nu_3=\nu > 1;\:\ \nu_2=1$): \; 
$\sigma=\nu_1+\nu+1$ \newline
This is the less singular case (${\cal O}(1/\e)$)
\beqa
&& I_{1m}^{IR}(D=2(\nu_1+\nu-\e); \nu_1,\nu,1)
= -\cG\frac{(-1)^{\nu_1+\nu}}{(\nu_1+\nu-2)!}\frac{1}{(\nu-1)}\frac{1}{\pt^2}
\nonumber \\ 
&\times&
\Bigg[\frac{1}{\e}-\ln(-\pt^2)-H_1(\nu-1)+2H_1(\nu_1+\nu-1)\Bigg]+{\cal O}(\e)\,.
\nonumber \\ 
\eeqa
\item $\nu_2>1; \nu_3>1$:\ $\sigma=\nu_1+\nu_2+\nu_3$ \newline
This is the finite case (${\cal O}(1)$)
\beqa
I_{1m}^{IR}(D=2(\sigma-1-\e); \nu_1,\nu_2,\nu_3)
&=& \frac{\cG (-1)^{\nu_1+\nu_2+\nu_3+1}}{\pt^2(\nu_1+\nu_2+\nu_3-3)! (\nu_2-1)(\nu_3-1)}
\nonumber \\  
&&+{\cal O}(\e)\,.
\eeqa
\end{enumerate}
\item {\bf The UV case}: $n \geq 0$ \newline
In this case we simply get from eq.~(\ref{basictriangle})
\beqa \label{duv}
I_{1m}^{UV}(D=2(\sigma-n-\e); \nu_1,\nu_2,\nu_3)&=&
\cG (-1)^{\nu_1+\nu_2+\nu_3}(\pt^2)^n
\frac{(\nu_2)_n(\nu_3)_n}{n!(\nu_1+\nu_2+\nu_3+2n-1)!}
\nonumber \\ && \times \left[
\frac{1}{\e}-\ln(-\pt^2)+H_1(n+1)+2 H_1(\nu_1+\nu_2+\nu_3+2n)
\right. \nonumber \\ && \left.
-H_1(\nu_2+n)-H_1(\nu_3+n)
\right]+{\cal O}(\e)\,.
\eeqa
\end{itemize}

\subsubsection{Two off-shell legs}
\beq
I_3(D=4-2\e;1,1,1 | \{\pu^2,\pd^2,0\})=\frac{\cG}{\epsilon^2} 
\frac{1}{\pu^2-\pd^2} \Big((-\pu^2)^{-\epsilon}-(-\pd^2)^{-\epsilon}\Big)\,.
\eeq
\subsubsection{Three off-shell legs}
The basis integral is the finite four dimensional 
triangle~\cite{Lu:1992ny,Usyukina:1992jd}
\beqa
\label{i3m}
I_3(D=4-2\e;1,1,1\, |\, \{\pu^2,\pd^2,\pt^2\})&=&
\frac{1}{\sqrt{\lambda(x,y)} \pt^2}
\Bigg( 2 \Big[\li(-\rho\, x) +\li(-\rho\, y)\Big] + \ln(\rho\, x) \ln(\rho\, y)
\nn
&+& 
 \ln\Big(\frac{x}{y}\Big) 
\ln\Big(\frac{1+\rho\, x}{1+ \rho\, y}\Big)+\frac{\pi^2}{3}  \Bigg)
+{\cal O}(\e)\,,
\eeqa
where $x=\pu^2/\pt^2,y=\pd^2/\pt^2$ and 
\beq
\lambda(x,y)={(1-x-y)^2-4 x y},\;\;\;
\rho(x,y)=\frac{2}{1-x-y+\sqrt{\lambda(x,y)}}\,. 
\eeq
Eq.~\eqref{i3m} is valid in the region where all $p_i^2 <0$ and
$\lambda(x,y) >0$. Details about the analytical continuation to other
regions can be found in reference~\cite{Bern:1997sc}.

\subsection{Box Integrals: $I_4(D=4-2\e;1,1,1,1)$}
\begin{figure}[t]
\includegraphics[angle=270,scale=0.55]{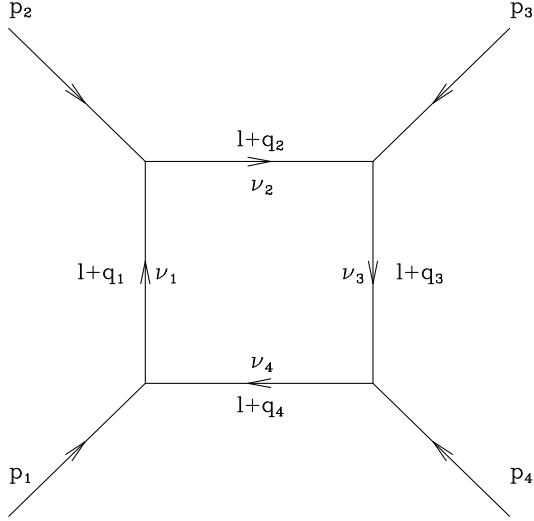}
\caption{The generic box diagram.}
\label{fig:boxdiag}
\end{figure}
All $4- 2 \epsilon$ dimensional box integrals with $\sigma=4$ belong in
the basis set.  The basic diagram is shown in Fig.~\ref{fig:boxdiag}.
The integrals are defined in terms of $s_{12}\equiv (\pu+\pd)^2 =
(\qd-\qq)^2$ and $s_{23} \equiv(\pd+\pt)^2 = (\qt-\qu)^2$ and the
off-shellness of the four legs, $p_i^2$.

We give here the explicit expression for the 4-dimensional boxes for
massless internal lines and any number of external masses.

\subsubsection{Four on-shell massless legs}
\vspace{-1cm}
\beqa
&&I_4(D=4-2\e;1,1,1,1,\{\sud,\sdt,0,0,0,0\})
=\frac{\cG}{\sud \sdt} \nonumber \\
&\times& \Big[\frac{2}{\e^2} \Big((-\sud)^{-\e}+(-\sdt)^{-\e}\Big)
-\ln^2\Big(\frac{-\sud}{-\sdt}\Big) - \pi^2 \Big]+{\cal O}(\e)\,.
\eeqa
\subsubsection{One off-shell leg}
\vspace{-1cm}
\beqa
&&I_4(D=4-2\e;1,1,1,1,\{\sud,\sdt,0,0,0,\pq^2\})
=\frac{\cG}{\sud \sdt} \nonumber \\
&\times& \Big[\frac{2}{\e^2} \Big((-\sud)^{-\e}+(-\sdt)^{-\e}-(-\pq^2)^{-\e}\Big)
 -2 \,\li(1-\frac{\pq^2}{\sud}) - 2 \, \li(1-\frac{\pq^2}{\sdt}) \nonumber \\
  &-& \ln^2\Big(\frac{-\sud}{-\sdt}\Big)-\frac{\pi^2}{3}\Big]+{\cal O}(\e)\,.
\eeqa
\subsubsection{Two contiguous off-shell legs}
\vspace{-1cm}
\beqa
&&I_4(D=4-2\e;1,1,1,1,\{\sud,\sdt,0,0,\pt^2,\pq^2 \})=
\frac{\cG}{\sud \sdt} \nonumber \\
  &\times& \Big[\frac{2}{\e^2} \Big((-\sud)^{-\e}+(-\sdt)^{-\e}
-(-\pt^2)^{-\e}-(-\pq^2)^{-\e})
  +\frac{1}{\e^2} \Big((-\pt^2)^{-\e} (-\pq^2)^{-\e}\Big)/(-\sud)^{-\e} \nonumber \\
  &-& 2\, \li\left(1-\frac{\pt^2}{ \sdt}\right)
     -2\, \li\left(1-\frac{\pq^2}{ \sdt}\right)
     -\ln^2\left(\frac{-\sud}{-\sdt}\right) \Big]+{\cal O}(\e)\,.
\eeqa
\subsubsection{Two opposite off-shell legs}
\vspace{-1cm}
\beqa
&&I_4(D=4-2\e;1,1,1,1,\{\sud,\sdt,0,\pd^2,0,\pq^2 \})
= \frac{\cG}{(\sdt\sud- \pq^2 \pd^2)} \nonumber \\
& \times &
\Big[\frac{2}{\e^2} 
\Big((-\sud)^{-\e}+(-\sdt)^{-\e}
      -(-\pd^2)^{-\e}-(-\pq^2)^{-\e}\Big) \nonumber \\
 &-&2\,\li\left(1-\frac{\pd^2}{ \sud}\right)                      
    -2\,\li\left(1-\frac{\pd^2}{ \sdt}\right)                      
    -2\,\li\left(1-\frac{\pq^2}{ \sud}\right)                      
    -2\,\li\left(1-\frac{\pq^2}{ \sdt}\right) \nonumber \\
    &+&2\, \li\left(1-\frac{\pd^2 \pq^2 }{ \sud\sdt}\right)
    -\ln^2\left(\frac{-\sud}{-\sdt}\right) \Big]+{\cal O}(\e)\,.
\eeqa
The pole at $\sdt\sud=\pq^2 \pd^2$ is only apparent
because it is canceled by the numerator.
In fact, setting $\pd^2 \pq^2=\sud\sdt (1-r)$ and performing the
expansion in $\e$ we find that the integral
may be approximated as
\beqa
&&I_4(D=4-2\e;1,1,1,1,\{\sud,\sdt,0,\pd^2,0,\pq^2 \})
= \frac{\cG}{\sdt\sud} \nonumber \\
& \times &
\Bigg\{ -\frac{1}{\e}(2+r)+2-\frac{r}{2}
 +(2+r) (\ln(-\sud)+\ln(-\sdt)-\ln(-\pq^2)) \nonumber \\
&+&2 \, \Ll_0\Big(\frac{-\pq^2}{-\sdt}\Big) 
+2 \, \Ll_0\Big(\frac{-\pq^2}{-\sud}\Big) 
    +r \Big[\Ll_1\Big(\frac{-\pq^2}{-\sdt}\Big)
         +\Ll_1\Big(\frac{-\pq^2}{-\sud}\Big) \Big]\Bigg\}+{\cal O}(\e,r^2)\,,
\eeqa
where $L_0, L_1$ are defined as,
\beq
\Ll_0(r) = \frac{\ln(r)}{1-r}\,,  \qquad 
\Ll_1(r) = \frac{\Ll_0(r)+1}{1-r} \; .
\eeq
The functions $L_0$ and $L_1$ have the property that they are finite as 
their denominators vanish.  

\subsubsection{Three off-shell legs}
The result for the box integral 
with three off-shell legs is \cite{Bern:1993kr}
\beqa
&&I_4(D=4-2\e;1,1,1,1,\{\sud,\sdt,0,\pd^2,\pt^2,\pq^2 \})=\frac{\cG}
{(\sdt\sud- \pd^2 \pq^2)}\nonumber \\
  &\times& \Big[\frac{2}{\e^2} \Big((-\sud)^{-\e}+(-\sdt)^{-\e}
-(-\pd^2)^{-\e}-(-\pt^2)^{-\e}-(-\pq^2)^{-\e}) \nonumber \\
  &+&\frac{1}{\e^2} \Big((-\pd^2)^{-\e} (-\pt^2)^{-\e}\Big)/(-\sdt)^{-\e} 
    +\frac{1}{\e^2} \Big((-\pt^2)^{-\e} (-\pq^2)^{-\e}\Big)/(-\sud)^{-\e} 
\\
  &-& 2\,\li\left(1-\frac{\pd^2}{ \sud}\right)
     -2\,\li\left(1-\frac{\pq^2}{ \sdt}\right)
     +2\,\li\left(1-\frac{\pd^2\pq^2}{\sud \sdt}\right)
-\ln^2 \Big(\frac{-\sud}{-\sdt}\Big) \Big]+{\cal O}(\e)\,.\nonumber
\eeqa

\subsubsection{Four off-shell legs}
In the case of four off-shell legs we have the remarkable
relation~\cite{Usyukina:1992jd}
\beq
I_4(D=4-2\e;1,1,1,1,\{\sud,\sdt,\pu^2,\pd^2,\pt^2,\pq^2 \})=
I_3(D=4-2\e;1,1,1,\{\sud \sdt,\pu^2\pt^2,\pd^2\pq^2\}), 
\eeq
which is valid for the case when all momenta are space-like. For the
continuation to other regions we refer to ref.~\cite{Denner:1991qq}.

\subsection{Pentagon Integrals: 
  $I_5(D=6-2\e;1,1,1,1,1)$} 
We choose as our final basis integral the finite six-dimensional
pentagon, $I_5(D=6-2\e; 1,1,1,1,1)$.  However at one loop the results
do not depend on the value of this finite integral, since it always
appears with a coefficient of
$\cO{\e}$~\cite{Bern:1993kr,Binoth:2005ff}.

\end{document}